\documentclass[11pt]{article}

\usepackage{graphicx,amsmath,amsfonts,amssymb,dcolumn,amsthm,euscript,braket}

\begin{document}

\title{\textbf{On a renormalizable  class of  gauge fixings for the  gauge invariant operator $A_{\min }^{2}$
}}
\author{\textbf{M.~A.~L.~Capri$^1$}\thanks{caprimarcio@gmail.com}, 
\textbf{D.~M. van Egmond$^1$} \thanks{duifjemaria@gmail.com}, 
\textbf{G. Peruzzo$^1$}\thanks{gperuzzofisica@gmail.com},
\textbf{M.~S.~Guimaraes$^1$}
\thanks{msguimaraes@uerj.br}, \\ 
\textbf{O. Holanda$^1$}\thanks{ozorio.neto@uerj.br},  
\textbf{S.~P.~Sorella$^1$}\thanks{silvio.sorella@gmail.com}, 
\textbf{R. C. Terin}$^{1}$\thanks{rodrigoterin3003@gmail.com}
\textbf{H. C. Toledo}$^{1}$\thanks{henriqcouto@gmail.com}\\\\\
\textit{{\small $^1$UERJ -- Universidade do Estado do Rio de Janeiro,}}\\
\textit{{\small Instituto de F\'isica -- Departamento de F\'{\i}sica Te\'orica -- Rua S\~ao Francisco Xavier 524,}}\\
\textit{{\small 20550-013, Maracan\~a, Rio de Janeiro, Brasil}}}
\date{}
\maketitle

\abstract{The dimension two gauge invariant non-local operator  $A_{\min }^{2}$, obtained through the minimization of 
$\int d^4x A^2$ along the gauge orbit, allows to introduce a non-local gauge invariant configuration $A^h_\mu$ which can be employed 
to built up a class of Euclidean massive Yang-Mills models useful to investigate non-perturbative infrared effects of confining theories. 
A fully local setup for both $A_{\min }^{2}$ and $A^{h}_\mu$ can be achieved, resulting in a local and BRST invariant action which 
shares similarities with the Stueckelberg formalism. Though, unlike the case of the Stueckelberg action, the use of $A_{\min }^{2}$ gives rise to an all orders 
renormalizable action, a feature which will be illustrated by means of a class of covariant gauge fixings which, as much as 't Hooft's $R_\zeta$-gauge of spontaneously broken gauge theories, provide a mass for 
the Stueckelberg field. }

\maketitle

\section{Introduction} 
In this work we pursue the investigation \cite{Fiorentini:2016rwx} of the dimension two gauge invariant operator $A_{\min }^{2}$, obtained by minimizing the functional 
$\mathrm{Tr}\int d^{4}x\,A_{\mu }^{u}A_{\mu }^{u}$ along the gauge
orbit of $A_{\mu }$ \cite{Zwanziger:1990tn,Dell'Antonio:1989jn,Dell'Antonio:1991xt,vanBaal:1991zw}, namely
\begin{eqnarray}
A_{\min }^{2} &\equiv &\min_{\{u\}}\mathrm{Tr}\int d^{4}x\,A_{\mu
}^{u}A_{\mu }^{u}\;,
\nonumber \\
A_{\mu }^{u} &=&u^{\dagger }A_{\mu }u+\frac{i}{g}u^{\dagger }\partial _{\mu
}u\;.    \label{Aminn0}
\end{eqnarray}
As highlighted in \cite{Fiorentini:2016rwx}, the functional $A_{\min }^{2}$ enables us to introduce a non-local gauge invariant field configuration $A^h_\mu$ \cite{Lavelle:1995ty}  which turns out to be helpful to construct renormalizable BRST invariant Yang-Mills theories which can be employed as effective massive theories to study non-perturbative
infrared aspects of confining Yang-Mills theories in Euclidean space. An example of such theories is provided by the so-called Refined Gribov-Zwanziger theory \cite{Dudal:2007cw,Dudal:2008sp,Dudal:2011gd}, recently formulated in a BRST invariant fashion \cite{Capri:2015ixa,Capri:2015nzw,Capri:2016aqq,Capri:2016gut,Capri:2017bfd}. We may also quote the massive model discussed in \cite{Tissier:2010ts,Tissier:2011ey}, where the mass term can be seen as to be deeply related to $A_{\min }^{2}$, thanks to the use of the Landau gauge, see eq.\eqref{land} below.  It is worth mentioning that a massive like behavior for the gluon propagator, known as the decoupling solution, has emerged within other approaches, as the study of the Schwinger-Dyson equations, the Renornalization Group  and other techniques, see for instance \cite{Aguilar:2008xm,Aguilar:2015bud,Fischer:2008uz,Aguilar:2015nqa,Huber:2015ria,Fischer:2009tn,Weber:2011nw,Frasca:2007uz,Siringo:2015wtx} and references therein.   The analytic decoupling solution for the two-point gluon correlation function turns out to be in very good agreement with the recent numerical simulations obtained on large lattices, both in Landau and linear covariant gauges \cite{Cucchieri:2007md,Cucchieri:2007rg,Cucchieri:2011ig,Oliveira:2012eh,Cucchieri:2009kk,Cucchieri:2011aa,Bicudo:2015rma,Cucchieri:2016jwg}.  \\\\In the present paper we extend the analysis of the operator $A_{\min }^{2}$ to a general class of covariant gauges which share  great similarity with 't Hooft's $R_\zeta$-gauge commonly used in the analysis of Yang-Mills theory with spontaneous symmetry breaking through the Higgs mechanism. In fact, as shown in \cite{Fiorentini:2016rwx}, the localization procedure for both $A_{\min }^{2}$ and $A^h_\mu$ requires the introduction of a dimensionless auxiliary Stueckelberg field $\xi$ which, as much as the Higgs field of 't Hooft's $R_\zeta$-gauge, will now enter explicitly the gauge condition through the appearance of a gauge massive paramater $\mu^2$.  This property will enable us to provide a fully BRST invariant mass for the auxiliary field $\xi$, a feature which might have helpful consequences in explicit loop calculations involving  $\xi$ in order to keep control of potential infrared divergences associated to its dimensionless nature.   Moreover, as in the case of $R_\zeta$-gauge, also the Faddeev-Popov ghosts will acquire a mass through the gauge-fixing. Of course, setting $\mu^2=0$, the linear covariant gauges discussed in \cite{Fiorentini:2016rwx} will be recovered. \\\\The work is organized as follows. In Sect.2 we give a short presentation of the main properties of $A_{\min }^{2}$ and of the gauge invariant configuration $A^h_\mu$, reminding to Appendix \eqref{apb} for more specific details. Sections 3,4,5,6 are devoted to the presentation of the local BRST invariant action for  $A_{\min }^{2}$ and $A^h_\mu$ as well as of the main properties of the aforementioned gauge-fixing. In Sect.7 we establish the set of Ward identities fulfilled by the resulting action. These identities will be employed to characterize the most general allowed invariant counterterm through the procedure of the algebraic renormalization  \cite{Piguet:1995er}. In Sect.8 a detailed analysis of the counterterm will be presented together with the renormalization factors needed to establish the all order renormalizability of the model. Section 9 contains our conclusion. Finally, in Appendix \eqref{apc},  a second, equivalent, proof of the renormalizability of the model will be outlined by making use of a generalised gauge fixing and ensuing Ward identities.

\section{Brief review of the operator $A_{\min }^{2}$ and construction of a non-local gauge invariant and transverse gauge field $A^h_\mu$}
For the benefit of the reader, let us give here a short overview of the operator $A_{\min }^{2}$, eq.\eqref{Aminn0}, reminding to the more complete Appendix \ref{apb} for details. \\\\In particular, looking at the the stationary condition for the functional \eqref{Aminn0},  one gets a non-local transverse
field configuration $A^h_\mu$, $\partial_\mu A^h_\mu=0$, which can be expressed as an infinite series in
the gauge field $A_\mu$, see  Appendix \ref{apb}, {\it i.e.}
\begin{eqnarray}
A_{\mu }^{h} &=&\left( \delta _{\mu \nu }-\frac{\partial _{\mu }\partial
_{\nu }}{\partial ^{2}}\right) \phi _{\nu }\;,  \qquad  \partial_\mu A^h_\mu= 0 \;, \nonumber \\
\phi _{\nu } &=&A_{\nu }-ig\left[ \frac{1}{\partial ^{2}}\partial A,A_{\nu
}\right] +\frac{ig}{2}\left[ \frac{1}{\partial ^{2}}\partial A,\partial
_{\nu }\frac{1}{\partial ^{2}}\partial A\right] +O(A^{3})\;.  \label{min0}
\end{eqnarray}
Remarkably, as shown in Appendix \ref{apb}, the configuration $A_{\mu }^{h}$ turns out to be left invariant
by infinitesimal gauge transformations order by order in the gauge
coupling $g$ \cite{Lavelle:1995ty}:
\begin{eqnarray}
\delta A_{\mu }^{h} &=&0\;,  \nonumber \\
\delta A_{\mu } &=&-\partial _{\mu }\omega +ig\left[ A_{\mu },\omega \right]
\;.  \label{gio}
\end{eqnarray}
Making use of \eqref{min0}, the gauge invariant nature of expression \eqref{Aminn0} can be made manifest by rewriting it in terms of the  field
strength $F_{\mu \nu }$. In fact, as proven in
\cite{Zwanziger:1990tn}, it turns out that
\begin{eqnarray}
A_{\min }^{2} = \int d^4x A_{\mu }^{h} A_{\mu }^{h} &=&-\frac{1}{2}\mathrm{Tr}\int d^{4}x\left( F_{\mu
\nu }\frac{1}{D^{2}}F_{\mu \nu }+2i\frac{1}{D^{2}}F_{\lambda \mu
}\left[ \frac{1}{D^{2}}D_{\kappa }F_{\kappa \lambda
},\frac{1}{D^{2}}D_{\nu }F_{\nu \mu }\right] \right.
\nonumber \\
&&-2i\left. \frac{1}{D^{2}}F_{\lambda \mu }\left[ \frac{1}{D^{2}}D_{\kappa
}F_{\kappa \nu },\frac{1}{D^{2}}D_{\nu }F_{\lambda \mu }\right] \right)
+O(F^{4})\;,  \label{zzw}
\end{eqnarray}
from which the gauge invariance becomes apparent. The operator $({D^{2}})^{-1}$ in expression 
\eqref{zzw} denotes the inverse of the Laplacian $D^2=D_\mu D_\mu$ with $D_\mu$ being the 
covariant derivative \cite{Zwanziger:1990tn}. Let us also underline that, in the Landau gauge 
$\partial_\mu A_\mu=0$, the operator $(A^h_\mu A^h_\mu)$ reduces to the operator $A^{2}$
\begin{equation}
(A^{h,a}_\mu A^{h,a}_\mu) \Big|_{\rm Landau} =  A^a_\mu A^a_\mu \;.      \label{land}
\end{equation}
This feature, combined with the gauge invariant nature of $(A^{h,a}_\mu A^{h,a}_\mu)$, implies that the 
anomalous dimension of $(A^{h,a}_\mu A^{h,a}_\mu)$ equals \cite{Fiorentini:2016rwx}, to all orders, that of the operator $A_{\mu }^{a}A_{\mu }^{a}$ of the Landau gauge, {\it i.e.} 
\begin{equation} 
\gamma_{(A^{h})^2} = \gamma_{A^{2}}\Big|_{\rm Landau} \;. \label{lld}
\end{equation} 
Moreover, as proven in \cite{Dudal:2002pq},  $\gamma_{A^{2}}\Big|_{\rm Landau}$ is not
an independent parameter, being given by 
\begin{equation}
\gamma_{A^{2}}\Big|_{\rm Landau} = \left(  \frac{\beta(a)}{a} + \gamma^{\rm Landau}_{A}(a)   \right)  \;, \qquad a = \frac{g^2}{16\pi^2}    \;,  \label{adl}
\end{equation}
where $(\beta(a), \gamma^{\rm Landau}_A(a))$ denote, respectively, the $\beta$-function and 
the anomalous dimension of the gauge field $A_\mu$ in the Landau gauge. This relation was conjectured and explicitly verified up to three-loop order in \cite{Gracey:2002yt}.

%%%%%%%%%%%%%%%%%%%%%%%%%%%%%%%%%%%%%%%%%%%%%%%%%%%%%
\section{A local action for   $A^h_\mu$}
\label{local_framework}
%%%%%%%%%%%%%%%%%%%%%%%%%%%%%%%%%%%%%%%%%%%%%%%%%%%%%
Following \cite{Fiorentini:2016rwx},  a fully local framework for the gauge invariant operator $A_{\mu }^{h}$ can be achieved. To that end, we consider the local, BRST invariant, action
 \begin{equation}
S_{inv} = \int d^{4}x\,\frac{1}{4}\,F^{a}_{\mu\nu}F^{a}_{\mu\nu} +
\int d^4x \left(\tau^{a}\,\partial_{\mu}A^{h,a}_{\mu}
+\frac{m^{2}}{2}\,A^{h,a}_{\mu}A^{h,a}_{\mu} + {\bar \eta}^a\partial_{\mu}D^{ab}_{\mu}(A^h)\eta^{b} \right)  \;,  \label{act1}
\end{equation}
where
\begin{equation}\label{local_Ah}
A^{h}_{\mu} \equiv A^{h,a}_{\mu}\,T^{a}=h^{\dagger}A_{\mu}h+\frac{i}{g}h^{\dagger}\partial_{\mu}h.
\end{equation}
with
\begin{equation}
h=e^{ig\xi}=e^{ig\xi^{a}T^{a}}.     \label{hxi}
\end{equation}
The matrices $\{T^a\}$ are the generators of the gauge group $SU(N)$ and $\xi^{a}$ is an auxiliary localizing Stueckelberg field. \\\\By expanding
(\ref{local_Ah}), one finds an infinite series whose first terms are
\begin{equation}\label{Ah_expansion}
 (A^{h})^{a}_{\mu}=A^{a}_{\mu}- \partial_{\mu} \xi^a - gf^{abc} A^b_\mu \xi^c -\frac{g}{2}f^{abc}\xi^{b}\partial_\mu \xi^c
 + {\rm higher \; orders}\,.
 \end{equation}
That the action $S_{inv}$ gives a local setup for the nonlocal operator $A^h_\mu$ of eq.\eqref{min0} follows by noticing that the Lagrange multiplier $\tau$ implements precisely the transversality condition
 \begin{equation}
 \partial_\mu A^h_\mu = 0  \;, \label{tr}
 \end{equation}
 which, when solved iteratively for the Stueckelberg field $\xi^a$, gives back  the expression \eqref{min0}, see  Appendix \ref{apb}. In addition, the extra ghosts $({\bar \eta}, \eta)$ account for the Jacobian arising from the functional integration over $\tau$ which gives a delta-function of the type  $\delta(\partial A^h)$. Finally, the term $\frac{m^{2}}{2}\,A^{h,a}_{\mu}A^{h,a}_{\mu}$  accounts for the inclusion of the gauge invariant operator  $A^{h,a}_{\mu}A^{h,a}_{\mu}$ through the mass parameter $m^2$ which, as mentioned before, can be used as an effective infrared parameter whose value can be estimated through comparison with the available lattice simulations on the two-point gluon correlation function, see \cite{Dudal:2007cw,Dudal:2008sp,Dudal:2011gd,Tissier:2010ts,Tissier:2011ey,Aguilar:2008xm,Aguilar:2015bud,Fischer:2008uz,Aguilar:2015nqa,Huber:2015ria,Fischer:2009tn,Weber:2011nw,Frasca:2007uz,Siringo:2015wtx,Cucchieri:2007md,Cucchieri:2007rg,Cucchieri:2011ig,Oliveira:2012eh,Cucchieri:2009kk,Cucchieri:2011aa,Bicudo:2015rma,Cucchieri:2016jwg} \\\\The local action $S_{inv}$, eq.\eqref{act1}, enjoys an exact BRST symmetry:
\begin{equation}
s S_{inv} = 0 \;, \label{brst_s}
\end{equation}
where the nilpotent BRST transformations are given by
\begin{eqnarray}
sA^{a}_{\mu}&=&-D^{ab}_{\mu}c^{b}\,,\nonumber \\
sc^{a}&=&\frac{g}{2}f^{abc}c^{b}c^{c}\,, \nonumber \\
s\bar{c}^{a}&=&ib^{a}\,,\nonumber \\
sb^{a}&=&0\,, \nonumber \\
s \tau^a & = & 0\,, \nonumber \\
s {\bar \eta}^a & = & s \eta^a = 0 \,, \nonumber \\
s^2 &=0 &\;.    \label{brst}
\end{eqnarray}
For the Stueckelberg field one has \cite{Dragon:1996tk},  with $i,j$ indices associated with a generic representation,
\begin{equation}
s h^{ij} = -ig c^a (T^a)^{ik} h^{kj}  \;, \qquad s (A^h)^a_\mu = 0  \;,  \label{brstst}
\end{equation}
from which the BRST transformation of the field $\xi^a$ can be evaluated iteratively, yielding
\begin{equation}
s \xi^a=  g^{ab}(\xi) c^b = - c^a + \frac{g}{2} f^{abc}c^b \xi^c - \frac{g^2}{12} f^{amr} f^{mpq} c^p \xi^q \xi^r + O(\xi^3)    \;.
\label{eqsxi}
\end{equation}

\section{Introducing the gauge fixing term $S_{gf}$}
As it stands, the action  \eqref{act1} needs to be equipped with the gauge fixing term, $S_{gf}$, which we choose as 
\begin{eqnarray}
S_{gf} &=&  \int d^4x \;s \left( {\bar c}^a (\partial_\mu A^a_\mu - \mu^2 \xi^a ) - {i} \frac{\alpha}{2} {\bar c^a} b^a \right)  \nonumber \\ 
&=&  \int d^4x \left( i b^a \partial_\mu A^a_\mu +  \frac{\alpha}{2} b^a b^a 
- i \mu^2 b^a \xi^a + {\bar c}^a \partial_\mu D^{ab}_\mu(A)c^b + \mu^2 {\bar c}^a g^{ab}(\xi) c^b  \right)    \;. \label{gf1}
\end{eqnarray}
Besides the traditional gauge parameter $\alpha$, we have now introduced a second gauge massive parameter $\mu^2$. As it will be clear in the next section, this massive parameter will provide a fully BRST invariant regularizing mass for the Stueckelberg field $\xi^a$, a feature which has helpful consequences when performing explicit loop calculations involving $\xi^a$. Setting $\mu^2=0$, the gauge fixing  \eqref{gf1} reduces to that of the usual linear covariant gauge \cite{Fiorentini:2016rwx}. Moreover, when $\mu^2=\alpha=0$, the Landau gauge, $\partial_\mu A^a_\mu=0$, is recovered. Nevertheless, it is worth underlining that both $\mu^2$ and $\alpha$ appear only in the gauge fixing term, which is an exact BRST variation. As such, $\mu^2$ and $\alpha$ are pure gauge parameters which will not affect the correlation functions of local BRST invariant operators. \\\\Though, before going any further, let us provide a few remarks related to the explicit presence  of the Stueckelberg field $\xi^a$ in eq.\eqref{gf1}. As it is easily realized, the field $\xi^a$ is a dimensionless field, a feature encoded in the fact that the invariant action $S_{inv}$ itself is an infinite series in powers of $\xi^a$. As in any local quantum field theory involving dimensionless fields, one has the freedom of performing arbitrary reparametrization of these fields as, for instance, in the case of  the two-dimensional non-linear sigma model \cite{Blasi:1988sh,Becchi:1988nh} and of  $N=1$ super Yang-Mills in superspace  \cite{Piguet:1981fb,Piguet:1981hh}. In the present case, this means that we have the freedom of replacing $\xi^a$ by an arbitrary dimensionless function of $\xi^a$, namely
\begin{equation} 
\xi^{a} \rightarrow \omega^a(\xi) = \xi^a + a_1^{abc} \xi^b \xi^c + a_2^{abcd} \xi^b \xi^c \xi^d + a_3^{abcde} \xi^b \xi^c \xi^d \xi^e + ........   \label{rp} 
\end{equation}    
This freedom, inherent to the dimensionless nature of $\xi^a$, is clearly evidentiated at the quantum level by the fact that the Stueckelberg field renormalizes in a non-linear way  \cite{Fiorentini:2016rwx}, {\it i.e.}  like eq.\eqref{rp}, expressing precisely the freedom one has in the choice of a reparametrization for $\xi^a$.  \\\\In our context, in eq.\eqref{gf1}, we could have been equally started  with a term like 
\begin{equation}
  s \left( {\bar c}^a  \xi^a \right) \rightarrow      s \left( {\bar c}^a  \omega^a(\xi)  \right) =  s \left( {\bar c}^a  ( \xi^a + a_1^{abc} \xi^b \xi^c + a_2^{abcd} \xi^b \xi^c \xi^d +...)  \right)  \;. \label{rp1}
\end{equation}
Of course, as much as $\mu^2$ and $\alpha$, all coefficients $(a_1^{abc}, a_2^{abcd}, a_3^{abcde}, ...)$ are gauge parameters, not affecting the correlation functions of the gauge invariant quantities. Equation \eqref{rp1} expresses the freedom which one always has when dealing with a gauge fixing term which depends explicitly from a dimensionless field, as the term \eqref{gf1}. In particular, this freedom will persist through the renormalization analysis, meaning that the renormalization of the gauge fixing itself has to be determined modulo an exact BRST terms of the kind $s\left( {\bar c}^a  \omega^a(\xi)  \right)$. Alternatively, one could start directly with the generalized gauge-fixing 
\begin{eqnarray}
S_{gf}^{gen} &=&  \int d^4x \;s \left( {\bar c}^a (\partial_\mu A^a_\mu) - \mu^2 \omega^a(\xi) ) - {i} \frac{\alpha}{2} {\bar c^a} b^a \right)  \nonumber \\ 
&=&  \int d^4x \left( i b^a \partial_\mu A^a_\mu +  \frac{\alpha}{2} b^a b^a 
- i \mu^2 b^a \omega^a(\xi) + {\bar c}^a \partial_\mu D^{ab}_\mu(A)c^b + \mu^2 {\bar c}^a \frac{\partial \omega^{a}(\xi)}{\partial \xi^c} g^{cd}(\xi) c^d  \right)    \;, \nonumber \\
\label{gf12}
\end{eqnarray}
and take into account the  renormalization of the quantity $\omega^a(\xi)$, encoded in the infinte set of gauge parameters $(a_1^{abc}, a_2^{abcd}, a_3^{abcde}, ...)$. In the following, we shall make use of the gauge-fixing \eqref{gf1} and identify in the final counterterm the term which corresponds to the reparametrization \eqref{rp1}. Moreover, in the Appendix \ref{apc}, we shall provide a second proof of the renormalizability of the model by deriving the generalized Slavnov-Taylor identities corresponding to the gauge fixing term \eqref{gf12}. \\\\In summary, as starting point, we shall take  the local, BRST invariant action 
\begin{equation}
S = S_{inv} +  S_{gf} \;, \label{stact}
\end{equation}
with 
\begin{equation}
s S = 0\;, \label{sinv}
\end{equation}
where the BRST transformations are given by eqs.\eqref{brst},\eqref{brstst},\eqref{eqsxi}.  \\\\Let us proceed now by giving a look at the propagators of the elementary fields. 

\section{A look at the propagators of the elementary fields} 

The propagators of the elementary fields are easily evaluated from the quadratic part of the action, eq.\eqref{stact}, {\it i.e.} 

\begin{eqnarray}
S_{quad.} & = & \int d^{4}x\; \left( \frac{1}{4} {(\partial_\mu A^a_\nu - \partial_\nu A^a_\mu)}^2 +ib^{a}\partial_{\mu}A_{\mu}^{a}+\frac{\alpha}{2}b^{a}b^{a}+\bar{c}^{a}\partial^{2}c^{a}-\mu^2 \bar{c}^{a}c^{a} \right.  \nonumber \\
 &  & +\frac{m^{2}}{2}A_{\mu}^{a}A_{\mu}^{a}-m^{2}A_{\mu}^{a}\partial_{\mu}\xi^{a}+\frac{m^{2}}{2}\left(\partial_{\mu}\xi^{a}\right)\left(\partial_{\mu}\xi^{a}\right) \nonumber \\
 &  & \Bigl.+\tau^{a}\partial_{\mu}A_{\mu}^{a}-\tau^{a}\partial^{2}\xi^{a}+\bar{\eta}^{a}\partial^{2}\eta^{a}-i\mu^2 b^{a}\xi^{a} \Bigr)  \nonumber \\
 & = & \int d^{4}x\; \frac{1}{2}\left[\begin{array}{cccc}
 A_{\mu}^{a} & b^{a} & \xi^{a} & \tau^{a}\end{array}\right] \times \nonumber \\
 &  & \times  \left[\begin{array}{cccc}
 \left(-\delta_{\mu\nu}\partial^{2}+\partial_{\mu}\partial_{\nu}+m^{2}\right) & -i\partial_{\mu} & -m^{2}\partial_{\mu} & -\partial_{\mu} \nonumber \\
i \partial_{\nu} & {\alpha} & - i\mu^{2} & 0 \nonumber \\
m^{2} \partial_{\nu} & -i\mu^{2} & -m^{2}\partial^{2} & -\partial^{2}\\
\partial_{\nu} & 0 & -\partial^{2} & 0
\end{array}\right] 
  \left[\begin{array}{c}
A_{\mu}^{a}\\
b^{a}\\
\xi^{a}\\
\tau^{a}
\end{array}\right]  \nonumber \\
&&+\int d^4x \; \left(\bar{c}^{a}\partial^{2}c^{a}+\bar{\eta}^{a}\partial^{2}\eta^{a}-\mu^{2}\bar{c}^{a}c^{a} \right)  \;.   \label{qqp} 
\end{eqnarray}
Thus, for the propagators we get
\begin{eqnarray}
\left\langle A_{\mu}^{a}\left(p\right)A_{\nu}^{b}\left(-p\right)\right\rangle  & = & \delta^{ab}\left(\frac{P_{\mu\nu}}{p^{2}+m^{2}}+\frac{\alpha p^{2}L_{\mu\nu}}{\left(p^{2}+\mu^{2}\right)^{2}}\right) \nonumber \\
\left\langle A_{\mu}^{a}\left(p\right)b^{b}\left(-p\right)\right\rangle  & = & \delta^{ab}\left(\frac{p_{\mu}}{p^{2}+\mu^{2}}\right) \nonumber \\
\left\langle A_{\mu}^{a}\left(p\right)\xi^{b}\left(-p\right)\right\rangle  & = & \delta^{ab}\left(\frac{-i\alpha p_{\mu}}{\left(p^{2}+\mu^{2}\right)^{2}}\right) \nonumber \\
\left\langle A_{\mu}^{a}\left(p\right)\tau^{b}\left(-p\right)\right\rangle  & = & \delta^{ab}\left(\frac{i\mu^{2}p_{\mu}}{p^{2}\left(p^{2}+\mu^{2}\right)}\right) \nonumber \\
\left\langle b^{a}\left(p\right)b^{b}\left(-p\right)\right\rangle  & = & 0 \nonumber \\
\left\langle b^{a}\left(p\right)\xi^{b}\left(-p\right)\right\rangle  & = & \frac{\delta^{ab}i}{p^{2}+\mu^{2}}\nonumber \\
\left\langle b^{a}\left(p\right)\tau^{b}\left(-p\right)\right\rangle  & = & 0 \nonumber \\
\left\langle \xi^{a}\left(p\right)\xi^{b}\left(-p\right)\right\rangle  & = & \frac{\delta^{ab}\alpha}{\left(p^{2}+\mu^{2}\right)^{2}} \nonumber \\
\left\langle \xi^{a}\left(p\right)\tau^{b}\left(-p\right)\right\rangle  & = & \frac{\delta^{ab}}{p^{2}+\mu^{2}} \nonumber \\
\left\langle \tau^{a}\left(p\right)\tau^{b}\left(-p\right)\right\rangle  & = & -\frac{\delta^{ab}m^{2}}{p^{2}} \nonumber \\
\left\langle {\bar c}^{a}\left(p\right)c^{b}\left(-p\right)\right\rangle  & = & \frac{\delta^{ab}}{p^{2}+\mu^2} \nonumber \\
\left\langle {\bar \eta}^{a}\left(p\right)\eta^{b}\left(-p\right)\right\rangle  & = & \frac{\delta^{ab}}{p^{2}} \label{propqq}
\end{eqnarray}
where, $P_{\mu\nu}=\left(\delta_{\mu\nu}-\frac{p_{\mu}p_{\nu}}{p^{2}}\right)$
and $L_{\mu\nu}=\frac{p_{\mu}p_{\nu}}{p^{2}}$ are the transverse and longitudinal projectors. We see that all propagators have a nice ultraviolet behavior, fully compatible with the power-counting. Moreover, the role of the massive gauge parameter $\mu^2$ becomes now apparent: it gives a BRST invariant regularizing mass for the Stueckelberg field $\xi^a$. Observe in fact that, when $\mu^2=0$, the propagator of the Stueckelberg field is given by $\langle \xi(p)\xi(-p)\rangle_{\mu^2=0}   =  \frac{\alpha}{p^{4}}$ which might give rise to potential infrared divergences in some class of Feynman diagrams. Notice also that, as expected, the mass parameter $m^2$ appears in the transverse part of the gluon propagator, a feature which exhibits its physical meaning. In fact, being coupled to the gauge invariant operator $(A^{h,a}_{\mu}A^{h,a}_{\mu})$, the parameter $m^2$ will enter the correlation functions of physical operators, {\it i.e.} gauge invariant operators, allowing thus to parametrize in an effective way their infrared behavior.

\section{$A_{\min }^{2}$ versus the conventional Stueckelberg mass term} \label{St} 
As done in \cite{Fiorentini:2016rwx}, before facing the analysis of the renormalizability of the action $S$, eq.\eqref{stact}, let us make a short comparison with the standard Stueckelberg mass term \cite{Ruegg:2003ps}, corresponding to the action 
\begin{equation} 
S_{Stueck} = \int d^{4}x\,\left(\frac{1}{4}\,F^{a}_{\mu\nu}F^{a}_{\mu\nu} +\frac{m^{2}}{2}\,A^{h,a}_{\mu}A^{h,a}_{\mu}  \right)  + S_{gf}  \;,  \label{stact1}
\end{equation}
where $S_{gf}$ is given by eq.\eqref{gf1}. One sees that the conventional Stueckelberg action corresponds to the addition of the gauge invariant operator $(A^{h,a}_{\mu}A^{h,a}_{\mu})$ without taking into account  the transversality constraint $\partial_\mu A^{h,a}_\mu=0$,  implemented in the action \eqref{stact} through the Lagrange multiplier $\tau^a$ and the corresponding ghosts $({\bar \eta}^a, \eta^a)$. The removal of the constraint $\partial_\mu A^{h,a}_\mu=0$ gives rise to the conventional Stueckelberg propagator, namely  
\begin{equation} 
\langle \xi^{a}\left(p\right)\xi^{b}\left(-p\right)\rangle_{Stueck}   =  \frac{\delta^{ab} p^2}{m^2 (p^2+\mu^2)^2} + \frac{\delta^{ab}\alpha}{\left(p^{2}+\mu^{2}\right)^{2}} \;. \label{stp}
\end{equation}
From this expression one easily understand the cause of the bad ultraviolet behavior of the Stueckelberg mass term, giving rise to its nonrenormalizability \cite{Ferrari:2004pd}.  We see in fact  that the mass parameter $m^2$ enters  the denominator of expression 
\eqref{stp}. As one easily figures out, this property jepardizes the renormalizability of the standard Stueckelberg formulation 
\cite{Ferrari:2004pd}. Due to the presence of the parameter $m^2$ in the denominator of expressions \eqref{stp}, non-renormalizable divergences in the inverse of the mass $m^2$ will show up, invalidating thus 
the perturbative loop expansion based on expression \eqref{stact1}. \\\\The role of the term  $\int d^4x\;\tau^{a}\,\partial_{\mu}A^{h,a}_{\mu}$, implementing the constraint $\partial_\mu A^{h,a}_\mu=0$, becomes now clear. It gives rise to a deep modification of the Stueckelberg propagator, removing precisely the first problematic term, $\frac{\delta^{ab} p^2}{m^2 (p^2+\mu^2)^2}$, from expression \eqref{stp}. We are left therefore only with the second piece, {\it i.e.}  $\frac{\delta^{ab}\alpha}{\left(p^{2}+\mu^{2}\right)^{2}}$, which does not pause any problem with the ultraviolet power-counting. It is this nice feature which will ensure the all order renormalizability of the action $S$, eq.\eqref{stact}, as we shall discuss in details in the next sections.

\section{Algebraic characterization  of the counterterm}\label{alg1} 
We are now ready to start the analysis of the renormalizability of the action  $S$, eq.\eqref{stact}. Following the setup of the algebraic renormalization \cite{Piguet:1995er}, we proceed by establishing the set of Ward identities which will be employed for the study of the quantum corrections. To that end, we need to introduce a set of external BRST invariant sources $(\Omega^a_\mu, L^a, K^a)$ coupled to the non-linear BRST variations of the fields $(A^a_\mu, c^a, \xi^{a})$ as well as sources $(\mathcal{J}_{\mu}^{a}, \Xi_{\mu}^{a})$ coupled to the BRST invariant  composite operators $( A_{\mu}^{ha},  D_{\mu}^{ab}(A^{h}))$, 
\begin{equation}
s \Omega^a_\mu = s L^a = sK^a = s \mathcal{J}_{\mu}^{a} = s \Xi_{\mu}^{a} = 0 \;.   \label{invs}
\end{equation} 
We shall thus start with the BRST invariant complete action $\Sigma$ defined by  
\begin{eqnarray}
\Sigma & = & \int d^{4}x\left( \frac{1}{4}\left(F_{\mu\nu}^{a}\right)^{2}+ib^{a}\partial_{\mu}A_{\mu}^{a}+\bar{c}^{a}\partial_{\mu}D_{\mu}^{ab}c^{b}+\frac{\alpha}{2}\left(b^{a}\right)^{2}-iM^{ab}b^{a}\xi^{b}\right. \nonumber \\
 &  & -N^{ab}\bar{c}^{a}\xi^{b}+M^{ab}\bar{c}^{a}g^{bc}\left(\xi\right)c^{c}+\bar{\eta}^{a}\partial_{\mu}D_{\mu}^{ab}\left(A^{h}\right)\eta^{b}+\frac{m^2}{2}A_{\mu}^{ha}A_{\mu}^{ha} \nonumber \\
 &  & +\tau^{a}\partial_{\mu}A_{\mu}^{ha}-\Omega_{\mu}^{a}D_{\mu}^{ab}c^{b}+\frac{g}{2}f^{abc}L^{a}c^{b}c^{c}+K^{a}g\left(\xi\right)^{ab}c^{b}+\mathcal{J}_{\mu}^{a}A_{\mu}^{ha} \nonumber \\
 &  & \Bigr.+\Xi_{\mu}^{a}D_{\mu}^{ab}\left(A^{h}\right)\eta^{b} \Bigl) \;, \label{cact}
\end{eqnarray}
where, for later convenience,  we have also introduced the BRST doublet of external sources $(M^{ab}, N^{ab})$ 
\begin{equation}
s M^{ab} = N^{ab} \;, \qquad s N^{ab} = 0  \;, 
\end{equation}
so that
\begin{equation}
s \Sigma = 0 \;. \label{Sg}
\end{equation} 
Notice that the invariant action $S$ of eq.\eqref{stact} is immediately recovered from the complete action $\Sigma$ upon setting the external sources  
$(\Omega_{\mu}^{a}=L^{a}=K^{a}=\mathcal{J}_{\mu}^{a}=\Xi_{\mu}^{a}=0)$ and $(M^{ab}=\delta^{ab}\mu^{2}, N^{ab}=0)$. \\\\It turns out that the complete action $\Sigma$ obeys the following Ward identities:  
 \begin{itemize}
\item the Slavnov-Taylor identity
\begin{eqnarray}
\mathcal{S}\left(\Sigma\right) & = & \int d^{4}x\left(\frac{\delta \Sigma}{\delta\Omega_{\mu}^{a}}\frac{\delta \Sigma}{\delta A_{\mu}^{a}}+\frac{\delta \Sigma}{\delta L^{a}}\frac{\delta \Sigma}{\delta c^{a}}+ib^{a}\frac{\delta \Sigma}{\delta\bar{c}^{a}}+\frac{\delta \Sigma}{\delta K^{a}}\frac{\delta \Sigma}{\delta\xi^{a}}+N^{ab}\frac{\delta \Sigma}{\delta M^{ab}}\right) = 0  \;,  \nonumber \\ \label{sti}
\end{eqnarray}
\item the equation of motion of the Lagrange multiplier  $b^{a}$ and of the antighost $\bar{c}^{a}$
\begin{eqnarray}
\frac{\delta \Sigma}{\delta b^{a}} & = & i\partial_{\mu}A_{\mu}^{a}+\alpha b^{a}-iM^{ab}\xi^{b} \;, \label{beq}
\end{eqnarray}
\begin{eqnarray}
\frac{\delta \Sigma}{\delta\bar{c}^{a}}+\partial_{\mu}\frac{\delta \Sigma}{\delta\Omega_{\mu}^{a}}-M^{ab}\frac{\delta \Sigma}{\delta K^{b}} & = & N^{ab}\xi^{b} \;, \label{agh}
\end{eqnarray}
\item the ghost-number Ward identity 
\begin{eqnarray}
\int d^{4}x\left(c^{a}\frac{\delta \Sigma}{\delta c^{a}}-\bar{c}^{a}\frac{\delta \Sigma}{\delta\bar{c}^{a}}-\Omega_{\mu}^{a}\frac{\delta \Sigma}{\delta\Omega_{\mu}^{a}}-2L^{a}\frac{\delta \Sigma}{\delta L^{a}}-K^{a}\frac{\delta \Sigma}{\delta K^{a}}+N^{ab}\frac{\delta \Sigma}{\delta N^{ab}}\right) & = & 0  \label{ghn}
\end{eqnarray}
\item the equation of the Lagrange multiplier $\tau^{a}$
\begin{eqnarray}
\frac{\delta \Sigma }{\delta\tau^{a}}-\partial_{\mu}\frac{\delta \Sigma}{\delta\mathcal{J}_{\mu}^{a}} & = & 0 \;, \label{teq}
\end{eqnarray}
\item the  $\eta^{a}$ Ward identity
\begin{eqnarray}
\int d^{4}x\left(\frac{\delta \Sigma}{\delta\eta^{a}}+gf^{abc}\bar{\eta}^{b}\frac{\delta \Sigma}{\delta\tau^{c}}+gf^{abc}\Xi^{b}\frac{\delta \Sigma}{\delta\mathcal{J}_{\mu}^{c}}\right) & = & 
0 \;, \label{etw}
\end{eqnarray}
\item the ${\bar \eta}^a$ antighost equation
\begin{eqnarray}
\frac{\delta \Sigma}{\delta\bar{\eta}^{a}}-\partial_{\mu}\frac{\delta \Sigma}{\delta\Xi_{\mu}^{a}} & = & 0 \;, \label{aetaw}
\end{eqnarray}
\item the $(\eta^a, \bar \eta^a)$ ghost number
\begin{eqnarray}
\int d^{4}x\left(\eta^{a}\frac{\delta \Sigma}{\delta\eta^{a}}-\bar{\eta}^{a}\frac{\delta \Sigma}{\delta\bar{\eta}^{a}}-\Xi^{a}\frac{\delta \Sigma}{\delta\Xi^{a}}\right) & = & 0 \;. \label{etew}
\end{eqnarray}
\end{itemize} 

\newpage 
 
\begin{table} 
\centering
\begin{tabular}{|c|c|c|c|c|c|c|c|c|c|c|c|c|c||c|c}
\hline 
 & $A_{\mu}^{a}$ & $b^{a}$ & $c^{a}$ & $\bar{c}^{a}$ & $\tau^{a}$ & $\eta^{a}$ & $\bar{\eta}^{a}$ & $\xi^{a}$  \tabularnewline
\hline 
\hline 
dim. & 1 & 2 & 0 & 2 & 2 & 0 & 2 & 0 \tabularnewline
\hline 
c gh. number  & 0 & 0 & 1 & -1 & 0 & 0 & 0 & 0 \tabularnewline
\hline 
$\eta$ gh. number  & 0 & 0 & 0 & 0 & 0 & 1 & -1 & 0 \tabularnewline
\hline    
\end{tabular}
\caption{The quantum numbers of the fields} 
 \label{tbb1}
 \end{table} 
\begin{table}
\centering
\begin{tabular}{|c|c|c|c|c|c|c|c|c|c|c|c|c|c||c|c}
\hline 
& $\Omega_{\mu}^{a}$ & $L^{a}$ & $K^{a}$ & $\mathcal{J}_{\mu}^{a}$ & $\Xi_{\mu}^{a}$ & $M^{ab}$ & $N^{ab}$\tabularnewline
\hline 
\hline 
dim. & 3 & 4 & 4 & 3 & 2 & 2 & 2\tabularnewline
\hline 
c gh. number   & -1 & -2 & -1 & 0 & 0 & 0 & 1\tabularnewline
\hline 
$\eta$ gh. number   & 0 & 0 & 0 & 0 & -1 & 0 & 0\tabularnewline
\hline    
\end{tabular}
\caption{The quantum numbers of the sources}
\label{tbb2} 
\end{table} 
\noindent All quantum numbers and dimensions of all fields and sources are displayed in  Tables \eqref{tbb1} and \eqref{tbb2}. \\\\In order to characterize the most general invariant counterterm  which can be freely added to all order
in perturbation theory, we follow the setup of the algebraic renormalization  \cite{Piguet:1995er} and 
perturb the classical action $\Sigma$, eq.\eqref{cact}, by adding an integrated local quantity in the fields and sources, 
$\Sigma^{ct}$, with dimension bounded by four and vanishing ghost number. We demand thus that the 
perturbed action, $(\Sigma +\varepsilon\Sigma^{ct})$, where $\varepsilon$ is an expansion parameter, 
fulfills, to the first order in $\varepsilon$, the same Ward identities obeyed by the classical action 
$\Sigma$, {\it i.e.} equations \eqref{sti}, \eqref{beq}, \eqref{ghn}, \eqref{teq}, \eqref{etw}  and \eqref{aetaw}. This amounts to impose the following 
constraints on $\Sigma$: 
 \begin{equation}
 {\cal B}_{\Sigma}  \Sigma^{ct} = 0 \;, \label{cc1} 
 \end{equation} 
 \begin{equation} 
 \frac{\delta \Sigma^{ct} }{\delta b^{a}}  = 0   \;, \label{cc2}
 \end{equation} 
 \begin{equation}
\frac{\delta \Sigma^{ct}}{\delta\bar{c}^{a}}+\partial_{\mu}\frac{\delta \Sigma^{ct}}{\delta\Omega_{\mu}^{a}}-M^{ab}\frac{\delta \Sigma^{ct}}{\delta K^{b}}  = 0   \;, \label{cc3}
\end{equation}
 \begin{equation}
\frac{\delta \Sigma^{ct} }{\delta\tau^{a}}-\partial_{\mu}\frac{\delta \Sigma^{ct}}{\delta\mathcal{J}_{\mu}^{a}}  =  0 \;,  \label{cc4}
\end{equation}
\begin{equation}
\int d^{4}x\left(\frac{\delta \Sigma^{ct}}{\delta\eta^{a}}+gf^{abc}\bar{\eta}^{b}\frac{\delta \Sigma^{ct}}{\delta\tau^{c}}+gf^{abc}\Xi^{b}\frac{\delta \Sigma^{ct}}{\delta\mathcal{J}_{\mu}^{c}}\right)  =  0 \;, \label{cc5}
 \end{equation} 
\begin{equation}
\frac{\delta \Sigma^{ct}}{\delta\bar{\eta}^{a}}-\partial_{\mu}\frac{\delta \Sigma^{ct}}{\delta\Xi_{\mu}^{a}}  =  0 \;, \label{cc6}
\end{equation} 
where  ${\cal B}_{\Sigma} $ is the so-called nilpotent linearized Slavnov-Taylor operator  \cite{Piguet:1995er}, defined as 
\begin{eqnarray}
{\cal B}_{\Sigma}  &= &  \int d^{4}x\left(\frac{\delta \Sigma}{\delta\Omega_{\mu}^{a}}\frac{\delta}{\delta A_{\mu}^{a}}+\frac{\delta \Sigma}{\delta A_{\mu}^{a}}\frac{\delta}{\delta\Omega_{\mu}^{a}}+\frac{\delta \Sigma}{\delta L^{a}}\frac{\delta}{\delta c^{a}}+\frac{\delta \Sigma}{\delta c^{a}}\frac{\delta}{\delta L^{a}}+\frac{\delta \Sigma}{\delta K^{a}}\frac{\delta}{\delta\xi^{a}} \right) \nonumber \\ 
&+& \int d^4x \left( \frac{\delta \Sigma}{\delta\xi^{a}}\frac{\delta}{\delta K^{a}}+ib^{a}\frac{\delta}{\delta\bar{c}^{a}}+N^{ab}\frac{\delta}{\delta M^{ab}}\right)   \;, \label{lst} 
\end{eqnarray}
with
\begin{equation} 
{\cal B}_{\Sigma} {\cal B}_{\Sigma}  =  0 \;.      \label{nplst}
\end{equation}
The first condition, eq.\eqref{cc1}, tells us that the counterterm $\Sigma^{ct}$ belongs to the cohomology of the operator ${\cal B}_{\Sigma} $ in the space of the integrated local polynomials in the fields, sources and parameters, of dimension four and ghost number zero. Owing to the general results on the BRST cohomolgy of Yang-Mills theories \cite{Piguet:1995er} and taking advantage of the analysis already done in \cite{Fiorentini:2016rwx}, the most general form for $\Sigma^{ct}$ can be written as  
\begin{eqnarray*}
\Sigma^{ct}  & = & \Delta_{cohom}+ {\cal B}_{\Sigma} \Delta^{\left(-1\right)} \;, \label{pct}
\end{eqnarray*}
where  $\Delta_{cohom}$ identifies the cohomolgy of ${\cal B}_{\Sigma} $, {\it i.e.} the non-trivial solution of eq.\eqref{cc1}, and  $\Delta^{\left(-1\right)}$ stands for the exact part, {\it i.e.} for the trivial solution of  \eqref{cc1}. Notice that, according to the quantum numbers of the fields, $\Delta^{\left(-1\right)}$ is an integrated polynomial of dimension four,  c-ghost number -1 and $\eta$-number equal to zero. \\\\For $\Delta_{cohom}$,  we have 
\begin{eqnarray}
\Delta_{cohom} & = & \int d^{4}x\; \Bigl( \frac{a_{0}}{4}\left(F_{\mu\nu}^{a}\right)^{2}+a_{1}\left(\partial_{\mu}A_{\mu}^{ha}\right)\left(\partial_{\nu}A_{\nu}^{ha}\right)+a_{2}\left(\partial_{\mu}A_{\nu}^{ha}\right)\left(\partial_{\mu}A_{\nu}^{ha}\right) \Bigr.  \nonumber \\
 &  & +a_{3}^{abcd}A_{\mu}^{ha}A_{\mu}^{hb}A_{\nu}^{hc}A_{\nu}^{hd}+\left(\partial_{\mu}\tau^{a}+\mathcal{J}_{\mu}^{a}\right)F_{\mu}^{a}\left(A,\xi\right)+a_{5}\left(\partial_{\mu}\bar{\eta}^{a}+\Xi_{\mu}^{a}\right)\left(\partial_{\mu}\eta^{a}\right) \nonumber \\
 &  & \Bigl. +f^{abc}\left(\partial_{\mu}\bar{\eta}^{a}+\Xi_{\mu}^{a}\right)\eta^{b}G_{\mu}^{c}\left(A,\xi\right)+m^{2}I\left(A,\xi\right)   \Bigr)   \;, \label{e1} 
\end{eqnarray}
where $F_{\mu}^{a}\left(A,\xi\right)$, $G_{\mu}^{c}\left(A,\xi\right)$
and $I\left(A,\xi\right)$ are local functional of $A_{\mu}^{a}$
and $\xi^{a}$, with dimension 1, 1 and 2, respectively. To write  expression  \eqref{e1} we have taken into account the constraints \eqref{cc3}--\eqref{cc6}. Moreover,  from condition \eqref{cc1} one immediately gets  
\begin{equation}
{\cal B}_{\Sigma} F_{\mu}^{a}\left(A,\xi\right)= {\cal B}_{\Sigma}G_{\mu}^{c}\left(A,\xi\right)={\cal B}_{\Sigma} I\left(A,\xi\right)= 0 \;. \label{t1}
\end{equation} 
Proceeding as in \cite{Fiorentini:2016rwx}, equations \eqref{t1} are solved by 
\begin{equation}
F_{\mu}^{a}\left(A,\xi\right) = a_4 A_{\mu}^{ha} \;, \qquad  G_{\mu}^{c}\left(A,\xi\right)= a_6  A_{\mu}^{ha} \;, \qquad  I\left(A,\xi\right)= a_7  A_{\mu}^{ha}A_{\mu}^{ha} \;, \label{t1a} 
\end{equation} 
where $(a_4, a_6, a_7)$ are free coefficients. Therefore, 
\begin{eqnarray*}
\Delta_{cohom} & = & \int d^{4}x\; \Bigl( \frac{a_{0}}{4}\left(F_{\mu\nu}^{a}\right)^{2}+a_{1}\left(\partial_{\mu}A_{\mu}^{ha}\right)\left(\partial_{\nu}A_{\nu}^{ha}\right)+a_{2}\left(\partial_{\mu}A_{\nu}^{ha}\right)\left(\partial_{\mu}A_{\nu}^{ha}\right) \Bigr. \nonumber \\
 &  & +a_{3}^{abcd}A_{\mu}^{ha}A_{\mu}^{hb}A_{\nu}^{hc}A_{\nu}^{hd}+a_{4}\left(\partial_{\mu}\tau^{a}+\mathcal{J}_{\mu}^{a}\right)A_{\mu}^{ha}+a_{5}\left(\partial_{\mu}\bar{\eta}^{a}+\Xi_{\mu}^{a}\right)\left(\partial_{\mu}\eta^{a}\right)\\
 &  & \Bigl. +a_{6}f^{abc}\left(\partial_{\mu}\bar{\eta}^{a}+\Xi_{\mu}^{a}\right)\eta^{b}A_{\mu}^{hc}+a_{7}m^{2}A_{\mu}^{ha}A_{\mu}^{ha} \Bigr) \;. \label{t1b}
\end{eqnarray*}
Let us discuss now the exact part of the cohomology of ${\cal B}_{\Sigma}$  which, taking into account the quantum numbers of the fields and sources,  can be parametrized as 
\begin{eqnarray*}
\Delta^{\left(-1\right)} & = & \int d^{4}x \; \Bigl( f_{1}^{ab}\left(\xi,\alpha\right)\xi^{a}K^{b}+f_{2}^{ab}\left(\xi,\alpha\right)L^{a}c^{b}+f_{3}^{ab}\left(\xi,\alpha\right)\xi^{a}\left(\partial_{\mu}\Omega_{\mu}^{b}\right)+f_{4}^{ab}\left(\xi,\alpha\right)\left(\partial_{\mu}\xi^{a}\right)\Omega_{\mu}^{b}. \Bigr. \nonumber \\
 &  & +f_{5}^{ab}\left(\xi,\alpha\right)A_{\mu}^{a}\Omega_{\mu}^{b}+f_{6}^{ab}\left(\xi,\alpha\right)A_{\mu}^{a}\left(\partial_{\mu}\bar{c}^{b}\right)+f_{7}^{ab}\left(\xi,\alpha\right)\left(\partial_{\mu}A_{\mu}^{a}\right)\bar{c}^{b}\\
 &  & +f_{8}^{ab}\left(\xi,\alpha\right)\left(\partial_{\mu}\xi^{a}\right)\left(\partial_{\mu}\bar{c}^{b}\right)+f_{9}^{ab}\left(\xi,\alpha\right)\xi^{a}\left(\partial^{2}\bar{c}^{b}\right)+f_{10}^{ab}\left(\xi,\alpha\right)\bar{c}^{a}b^{b}\\
 &  & +f_{11}^{ab}\left(\xi,\alpha\right)\bar{c}^{a}\tau^{b}+f_{12}^{abc}\left(\xi,\alpha\right)\bar{\eta}^{a}\eta^{b}\bar{c}^{c}+f_{13}^{abc}\left(\xi,\alpha\right)\bar{c}^{a}\bar{c}^{b}c^{c}\\
 &  & \Bigl. +f_{14}^{abcd}\left(\xi,\alpha\right)M^{ab}\xi^{c}\bar{c}^{d} \Bigr) \;, \label{trv1}
\end{eqnarray*}
where $(f_{1}, ..., f_{14})$ are arbitrary coefficients. Imposing the constraint \eqref{cc2}, {\it i.e.}
\begin{equation}
\frac{\delta}{\delta b^{k}} {\cal B}_{\Sigma} \Delta^{\left(-1\right)} = 0 \;, \label{bct} 
\end{equation} 
 and making use of the commutation relation 
\begin{equation}
\frac{\delta}{\delta b^{k}}{\cal B}_{\Sigma} ={\cal B}_{\Sigma} \frac{\delta}{\delta b^{k}}+i\left(\frac{\delta}{\delta\bar{c}^{k}}+\partial_{\mu}\frac{\delta}{\delta\Omega_{\mu}^{k}}-M^{kl}\frac{\delta}{\delta K^{l}}\right) \;, \label{cm}
\end{equation} 
one finds 
\begin{eqnarray*}
\frac{\delta\Delta^{\left(-1\right)}}{\delta b^{k}}  =  f_{10}^{ak}\left(\xi,\alpha\right)\bar{c}^{a}  \qquad 
\Rightarrow \qquad {\cal B}_{\Sigma}\frac{\delta\Delta^{\left(-1\right)}}{\delta b^{k}}  =  \frac{\delta \Sigma}{\delta K^{m}}\frac{\partial f_{10}^{ak}\left(\xi,\alpha\right)}{\partial\xi^{m}}\bar{c}^{a}+if_{10}^{ak}\left(\xi,\alpha\right)b^{a} \;. \label{cm1}
\end{eqnarray*}
Moreover, from 
\begin{eqnarray*}
i\left(\frac{\delta\Delta^{\left(-1\right)}}{\delta\bar{c}^{k}}+\partial_{\mu}\frac{\delta\Delta^{\left(-1\right)}}{\delta\Omega_{\mu}^{k}}-M^{kl}\frac{\delta\Delta^{\left(-1\right)}}{\delta K^{l}}\right) & = & -i\partial_{\mu}\left(f_{6}^{ak}\left(\xi,\alpha\right)A_{\mu}^{a}\right)+if_{7}^{ak}\left(\xi,\alpha\right)\left(\partial_{\mu}A_{\mu}^{a}\right)\\
 &  & -i\partial_{\mu}\left(f_{8}^{ak}\left(\xi,\alpha\right)\left(\partial_{\mu}\xi^{a}\right)\right)+i\partial^{2}\left(f_{9}^{ak}\left(\xi,\alpha\right)\xi^{a}\right)\\
 &  & +if_{10}^{kb}\left(\xi,\alpha\right)b^{b}+if_{11}^{kb}\left(\xi,\alpha\right)\tau^{b}\\
 &  & +if_{12}^{abk}\left(\xi,\alpha\right)\bar{\eta}^{a}\eta^{b}+2if_{13}^{kbc}\left(\xi,\alpha\right)\bar{c}^{b}c^{c}\\
 &  & +if_{14}^{abck}\left(\xi,\alpha\right)M^{ab}\xi^{c}\\
 &  & -i\partial^{2}\left(f_{3}^{ak}\left(\xi,\alpha\right)\xi^{a}\right)+i\partial_{\mu}\left(f_{4}^{ak}\left(\xi,\alpha\right)\left(\partial_{\mu}\xi^{a}\right)\right) \\
 &  & +i\partial_{\mu}\left(f_{5}^{ak}\left(\xi,\alpha\right)A_{\mu}^{a}\right)-iM^{kl}f_{1}^{al}\left(\xi,\alpha\right)\xi^{a} \;, \label{cm2}
\end{eqnarray*}
it follows that 
\begin{eqnarray*}
 \frac{\delta}{\delta b^{k}} {\cal B}_{\Sigma} \Delta^{\left(-1\right)}  = 0  & = & \left[\frac{\partial f_{10}^{bk}\left(\xi,\alpha\right)}{\partial\xi^{m}}g^{mc}\left(\xi\right)-2if_{13}^{kbc}\left(\xi,\alpha\right)\right]c^{c}\bar{c}^{b}\\
 &  & +i\left[f_{10}^{ak}\left(\xi,\alpha\right)+f_{10}^{ka}\left(\xi,\alpha\right)\right]b^{a}\\
 &  & +i\left[-f_{6}^{ak}\left(\xi,\alpha\right)+f_{5}^{ak}\left(\xi,\alpha\right)+f_{7}^{ak}\left(\xi,\alpha\right)\right]\left(\partial_{\mu}A_{\mu}^{a}\right)\\
 &  & -i\left[\left(\partial_{\mu}f_{6}^{ak}\left(\xi,\alpha\right)\right)-\left(\partial_{\mu}f_{5}^{ak}\left(\xi,\alpha\right)\right)\right]A_{\mu}^{a}\\
 &  & +i\left[-\left(\partial_{\mu}f_{8}^{ak}\left(\xi,\alpha\right)\right)-\left(\partial_{\mu}f_{3}^{ak}\left(\xi,\alpha\right)\right)+\left(\partial_{\mu}f_{4}^{ak}\left(\xi,\alpha\right)\right)+\left(\partial_{\mu}f_{9}^{ak}\left(\xi,\alpha\right)\right)\right]\left(\partial_{\mu}\xi^{a}\right)\\
 &  & +i\left[-f_{8}^{ak}\left(\xi,\alpha\right)-f_{3}^{ak}\left(\xi,\alpha\right)+f_{4}^{ak}\left(\xi,\alpha\right)+f_{9}^{ak}\left(\xi,\alpha\right)\right]\left(\partial^{2}\xi^{a}\right)\\
 &  & +i\left[-\left(\partial^{2}f_{3}^{ak}\left(\xi,\alpha\right)\right)+\left(\partial^{2}f_{9}^{ak}\left(\xi,\alpha\right)\right)\right]\xi^{a}\\
 &  & +if_{11}^{kb}\left(\xi,\alpha\right)\tau^{b}+if_{12}^{abk}\left(\xi,\alpha\right)\bar{\eta}^{a}\eta^{b}
    +i\left[f_{14}^{abck}\left(\xi,\alpha\right)-\delta^{ka}f_{1}^{cb}\left(\xi,\alpha\right)\right]M^{ab}\xi^{c} \;, \label{cm3}
\end{eqnarray*}
form which we can derive relations among the coefficients $(f_{1}, ..., f_{14})$.  Let us start with 
\begin{equation} 
\left(\partial_{\mu}f_{6}^{ak}\left(\xi,\alpha\right)\right)-\left(\partial_{\mu}f_{5}^{ak}\left(\xi,\alpha\right)\right)=0 \qquad \Rightarrow  \qquad  f_{6}^{ab}  =  f_{5}^{ab}+\delta^{ab}a \;, \label{cm4}
\end{equation} 
where $a$ is a constant. Further
\begin{equation} 
-f_{6}^{ak}\left(\xi,\alpha\right)+f_{5}^{ak}\left(\xi,\alpha\right)+f_{7}^{ak}\left(\xi,\alpha\right)=0 \qquad \Rightarrow \qquad 
f_{7}^{ak}\left(\xi,\alpha\right)=\delta^{ab}a \;. \label{cm5}
\end{equation} 
Analogously 
\begin{equation} 
-\left(\partial^{2}f_{3}^{ak}\left(\xi,\alpha\right)\right)+\left(\partial^{2}f_{9}^{ak}\left(\xi,\alpha\right)\right)=0 \qquad \Rightarrow \qquad f_{9}^{ak}\left(\xi,\alpha\right)=f_{3}^{ak}\left(\xi,\alpha\right)+b\delta^{ak} \;, \label{cm6}
\end{equation} 
with $b$ a free constant. Next, from 
\begin{equation} 
\left[-\left(\partial_{\mu}f_{8}^{ak}\left(\xi,\alpha\right)\right)-\left(\partial_{\mu}f_{3}^{ak}\left(\xi,\alpha\right)\right)+\left(\partial_{\mu}f_{4}^{ak}\left(\xi,\alpha\right)\right)+\left(\partial_{\mu}f_{9}^{ak}\left(\xi,\alpha\right)\right)\right]   \;, \label{cm7}
\end{equation} 
we get 
\begin{equation}
f_{8}^{ak}\left(\xi,\alpha\right)=f_{4}^{ak}\left(\xi,\alpha\right)+c\delta^{ak} \;, \label{cm8}
\end{equation} 
with $c$ constant. Finally 
 \begin{equation}
 -f_{8}^{ak}\left(\xi,\alpha\right)-f_{3}^{ak}\left(\xi,\alpha\right)+f_{4}^{ak}\left(\xi,\alpha\right)+f_{9}^{ak}\left(\xi,\alpha\right)=0 \qquad \Rightarrow \qquad b=c  \;.  \label{cm8} 
 \end{equation} 
 Therefore,  $\Delta^{\left(-1\right)}$ becomes 
\begin{eqnarray*}
\Delta^{\left(-1\right)} & = &
  \int d^{4}x \; \Bigl( f_{1}^{ab}\left(\xi,\alpha\right)\left(\xi^{a}K^{b}+M^{cb}\xi^{a}\bar{c}^{c}\right) \Bigr. \\
 &  & +f_{2}^{ab}\left(\xi,\alpha\right)L^{a}c^{b}+f_{3}^{ab}\left(\xi,\alpha\right)\xi^{a}\left(\left(\partial_{\mu}\Omega_{\mu}^{b}\right)+\left(\partial^{2}\bar{c}^{b}\right)\right)\\
 &  & +f_{4}^{ab}\left(\xi,\alpha\right)\left(\partial_{\mu}\xi^{a}\right)\left(\Omega_{\mu}^{b}+\left(\partial_{\mu}\bar{c}^{b}\right)\right)\\
 &  & +f_{5}^{ab}\left(\xi,\alpha\right)A_{\mu}^{a}\left(\Omega_{\mu}^{b}+\left(\partial_{\mu}\bar{c}^{b}\right)\right)\\
 &  & \Bigl. +f_{10}^{ab}\left(\xi,\alpha\right)\bar{c}^{a}b^{b}+\frac{1}{2i}\frac{\partial f_{10}^{ba}\left(\xi,\alpha\right)}{\partial\xi^{m}}g^{mc}\left(\xi\right)\bar{c}^{a}\bar{c}^{b}c^{c} \Bigr)  \;. \label{cm9} 
\end{eqnarray*}
We can now impose the constraint \eqref{cc5}  
\begin{eqnarray*}
\int d^{4}x\left(\frac{\delta \Sigma^{ct}}{\delta\eta^{m}}+gf^{mnp}\bar{\eta}^{n}\frac{\delta \Sigma^{ct}}{\delta\tau^{p}}+gf^{mnp}\Xi^{n}\frac{\delta \Sigma^{ct}}{\delta\mathcal{J}_{\mu}^{p}}\right) & = & 0 \;, \label{cm10} 
\end{eqnarray*}

\begin{eqnarray*}
\Rightarrow\int d^{4}x\left(a_{6}+a_{4}g\right)f^{mnp}\left(\partial_{\mu}\bar{\eta}^{n}+f^{mnp}\Xi^{n}\right)A_{\mu}^{hp} & = & 0 \;,  \label{cm11}
\end{eqnarray*}
from which we obtain $a_{6}=-a_{4}g$. \\\\As done in \cite{Fiorentini:2016rwx},  we can further reduce the number of parameters entering $\Sigma^{ct}$ by observing that, setting  
$K^{a}=M^{ab}=N^{ab}=\mathcal{J}_{\mu}^{a}=\Xi_{\mu}^{a}=m=0$, the complete action $\Sigma$, eq.\eqref{cact}, reduces to that or ordinary Yang-Mills theory in the linear covarinat gauges, as  integration over $\tau^{a}$, $\eta^{a}$ and $\overline{\eta}^{a}$
gives a unity. As a consequence, making use of the well known renormalization of standard Yang-Mills theory in the linear covariant gauges \cite{Piguet:1995er}, we get  $a_{1}=a_{2}=a_{3}^{abcd}=0$, $a_{5}=a_{4}$, as well as 

\begin{equation}
f_{2}^{ab}\left(\xi,\alpha\right)  =  \delta^{ab}d_{1}\left(\alpha\right) \;, \qquad  f_{5}^{ ab}\left(\xi,\alpha\right)  =  \delta^{ab}d_{2}\left(\alpha\right) \;, \label{cm12} 
\end{equation}
with $(d_{1}, d_{2})$ free parameters. In addition, we also have 
\begin{eqnarray*}
f_{3}^{ab}\left(\xi,\alpha\right)=f_{4}^{ab}\left(\xi,\alpha\right)=f_{10}^{ab}\left(\xi,\alpha\right) & = & 0 \;. \label{cm13} 
\end{eqnarray*}
Hence
\begin{eqnarray*}
\Delta_{cohom}   & = & \int d^{4}x\; \Bigl( \frac{a_{0}}{4}\left(F_{\mu\nu}^{a}\right)^{2}+a_{4}\left(\left(\partial_{\mu}\tau^{a}+\mathcal{J}_{\mu}^{a}\right)A_{\mu}^{ha}+\left(\partial_{\mu}\bar{\eta}^{a}+\Xi_{\mu}^{a}\right)D^{ab}\left(A^{h}\right)\eta^{b}\right)  \Bigr. \\
 &  & \Bigl. \; \; \; \; \; +a_{7}m^{2}A_{\mu}^{ha}A_{\mu}^{ha} \Bigr) \;, \label{cm14} 
\end{eqnarray*}
and

\begin{eqnarray*}
\Delta^{\left(-1\right)}  =  \int d^{4}x \Bigl( f_{1}^{ab}\left(\xi,\alpha\right)\left(\xi^{a}K^{b}+M^{cb}\xi^{a}\bar{c}^{c}\right) 
    +d_{1}\left(\alpha\right)L^{a}c^{a}+d_{2}\left(\alpha\right)A_{\mu}^{a}\left(\Omega_{\mu}^{a}+\left(\partial_{\mu}\bar{c}^{a}\right)\right) \Bigr) \;.  \label{cm15}
\end{eqnarray*}
Let us end this section by  rewriting the final expression of the most general invariant counterterm $\Sigma^{ct}$ in its parametric form \cite{Piguet:1995er}, a task that will simplify the analysis of the renormalziation factors, namely  

\begin{eqnarray}
\Sigma^{ct}  & = & -a_{0}g\frac{\partial \Sigma}{\partial g}+d_{2}\left(\alpha\right)2\alpha\frac{\partial \Sigma}{\partial\alpha}+a_{7}m^{2}\frac{\partial \Sigma}{\partial m^{2}} \nonumber \\
 &  & +\int d^{4}x\; \Bigl( a_{4}\left(-\tau^{a}\frac{\delta \Sigma}{\delta\tau^{a}}+\mathcal{J}_{\mu}^{a}\frac{\delta \Sigma}{\delta\mathcal{J}_{\mu}^{a}}-\bar{\eta}^{a}\frac{\delta \Sigma}{\delta\bar{\eta}^{a}}+\Xi_{\mu}^{a}\frac{\delta \Sigma}{\delta\Xi_{\mu}^{a}}\right) \Bigr.  \nonumber \\
 &  & -\left(f_{1}^{ab}\left(\xi,\alpha\right)+\frac{\partial f_{1}^{kb}\left(\xi,\alpha\right)}{\partial\xi^{a}}\xi^{k}\right)K^{b}\frac{\delta \Sigma}{\delta K^{a}}+f_{1}^{ab}\left(\xi,\alpha\right)\xi^{a}\frac{\delta \Sigma}{\delta\xi^{b}} \nonumber \\
 &  & +d_{2}\left(\alpha\right)A_{\mu}^{a}\frac{\delta \Sigma}{\delta A_{\mu}^{a}}-d_{2}\left(\alpha\right)b^{a}\frac{\delta \Sigma}{\delta b^{a}}
  -d_{2}\left(\alpha\right)\Omega_{\mu}^{a}\frac{\delta \Sigma}{\delta\Omega_{\mu}^{a}}-d_{2}\left(\alpha\right)\bar{c}^{a}\frac{\delta \Sigma}{\delta\overline{c}^{a}} \nonumber \\
 &  & -d_{1}\left(\alpha\right)c^{a}\frac{\delta \Sigma}{\delta c^{a}}+d_{1}\left(\alpha\right)L^{a}\frac{\delta \Sigma}{\delta L^{a}} +\left(-f_{1}^{cb}\left(\xi,\alpha\right)+d_{2}\left(\alpha\right)\delta^{cb}\right)N^{ab}\frac{\delta \Sigma}{\delta N^{ac}} \nonumber \\
 &  & \Bigl.+\left(d_{2}\left(\alpha\right)\delta^{ab}-f_{1}^{ab}\left(\xi,\alpha\right)\right)M^{cb}\frac{\delta \Sigma}{\delta M^{ca}} +\frac{\partial f_{1}^{ab}\left(\xi,\alpha\right)}{\partial\xi^{k}}M^{cb}\xi^{a}g\left(\xi\right)^{kd}c^{d}\bar{c}^{c}  \Bigr) \;. \label{ctf}
\end{eqnarray}

\section{Analysis of the counterterm and renormalization factors}\label{alg2} 
Having determined the most general form of the local invariant counterterm, eq.\eqref{ctf}, let us turn to its physical meaning. 
As already mentioned before, in order to determine the renormalization of the fields, sources and parameters, we have to 
pay attention to the fact that, due to the explicit dependence of the gauge fixing from the Stueckelberg field $\xi^a$,  the renormalization of the gauge fixing itself is determined up to an ambiguity of the type of eq.\eqref{rp1}, which would correspond to the renormalization of the quantity $\omega^a(\xi)$, {\it i.e.} of the gauge parameters $(a_1^{abc}, a_2^{abcd}, a_3^{abcde}, ...)$. To that end, it will be sufficient to analyse the last two terms of the expression for $\Sigma^{ct}$, eq.\eqref{ctf}, which, upon setting the sources $(M^{ab},N^{ab})$ to their physical values, namely $(M^{ab}= \delta^{ab} \mu^2, N^{ab}=0)$, becomes 
\begin{equation}
\left( d_2(\alpha) - f_1(0, \alpha) \right) \mu^2 \frac{\partial \Sigma}{\partial \mu^2}  + \mu^2 \int d^4x \left( {\tilde f}_1^{ab}(\xi, \alpha) (i b^b \xi^a - {\bar c}^b g^{ak}(\xi)c^k) + 
\frac{\partial {\tilde f}_1^{ab}(\xi, \alpha) }{\partial \xi^k} {\bar c}^b \xi^a g^{kd}(\xi) c^d \right)   \label{ct1}
\end{equation}
where we have set 
\begin{equation} 
f_1^{ab}(\xi, \alpha) = f_1^{ab}(0, \alpha) + {\tilde f}_1^{ab}(\xi, \alpha)  \;, \label{fd}  
\end{equation}
with $f_1^{ab}(0, \alpha) = \delta^{ab} f_1(0,\alpha)$ being the first, $\xi^a$-independent,  term of the Taylor expansion of $f_1^{ab}(\xi, \alpha)$ in powers of $\xi^a$ and $ {\tilde f}_1^{ab}(\xi, \alpha)$ denoting the $\xi$-dependent remaining terms. Of course, $f_1^{ab}(0, \alpha) = \delta^{ab} f_1(0,\alpha)$ is just a constant. \\\\Furthermore, we observe that expression \eqref{ct1} can be rewritten as 
\begin{equation}
\left( d_2(\alpha) - f_1(0, \alpha) \right) \mu^2 \frac{\partial \Sigma}{\partial \mu^2}  + \mu^2 \int d^4x \;s \left( {\tilde f}_1^{ab}(\xi,\alpha) {\bar c}^b \xi^a \right)  \;, \label{ct2} 
\end{equation}
or, equivalently 
\begin{equation}
\left( d_2(\alpha) - f_1(0, \alpha) \right) \mu^2 \frac{\partial \Sigma}{\partial \mu^2}  + \mu^2 \int d^4x \;s \left(  {\bar c}^b {\tilde \omega}^b(\xi, \alpha) \right) \;, \label{ct3}
\end{equation}
 with ${\tilde \omega}^b(\xi, \alpha) = {\tilde f}_1^{ab}(\xi,\alpha) \xi^a$. \\\\We are now able to unravel the meaning of this term. First, the term $\left( d_2(\alpha) - f_1^{aa}(0, \alpha) \right)$ corresponds to a multiplicative renormalization of the gauge massive parameter $\mu^2$. This follows by observing that, being $\mu^2$ a space-time independent parameter, its renormalization must be given by a field independent space-time constant factor, {\it i.e.} precisely by $\left( d_2(\alpha) - f_1^{aa}(0, \alpha) \right)$. On the other hand, the term $\int d^4x \;s \left(  {\bar c}^b {\tilde \omega}^b(\xi, \alpha) \right)$ is of the type of eq.\eqref{rp1}, thus corresponding to the ambiguity inherent to the gauge fixing discussed before. As already mentioned, this term can be handled by starting with the generalised gauge fixing  \eqref{gf12}, whose algebraic renormalization can be faced by employing the Ward identities displayed in Appendix \eqref{apc}. Doing so, the term $\int d^4x \;s \left(  {\bar c}^b {\tilde \omega}^b(\xi, \alpha) \right)$ will correspond to a renormalization of the gauge fixing function $\omega^a(\xi)$, {\it i.e.} of the gauge parameters $(a_1^{abc}, a_2^{abcd}, a_3^{abcde}, ...)$. \\\\We can now read off the renormalization factors, {\it i.e.} 
 \begin{equation} 
 \Sigma(\Phi) +  \varepsilon  \Sigma^{ct}(\Phi)   =   \Sigma(\Phi_0) + O(\varepsilon^2) \;, \label{ra1}
 \end{equation}
with
\begin{eqnarray}
\Phi_0= Z_{\Phi} \Phi  + O(\varepsilon^2) \;, \label{ra2}
\end{eqnarray}
where $\Phi$ stands for a short-hand notation for all fields, sources and parameters. Specifically, for the renormalization factors one finds:

\begin{eqnarray}
A_0&=&Z_A^{1/2} A_{\mu}\;,  \,\,\,  b_0=Z_b^{1/2}b\;,  \,\,\,  c_0=Z_c^{1/2}c \;, \,\,\, \bar{c}_0=Z_{\bar{c}}^{1/2}\bar{c} \;,  \\ 
\xi_0^a &= &Z^{ab}_{\xi}(\xi)\xi^b \;  \,\,\, \tau_0=Z_{\tau}^{1/2} \tau \;,  \Omega_0 =Z_{\Omega} \Omega \;, \,\, \,L_0=Z_L L \; \,\,\,  \\
K_0^a&=& Z_K^{ab} (\xi) K^b\;, \,\,\, m_0^2=Z_{m^2} m^2\;, \,\,\,\mathcal{J}_0=Z_{\mathcal{J}}\mathcal{J}\;,\\
g_0&=&Z_g g \;, \,\,\, \alpha_0=Z_{\alpha} \alpha \;, \,\,\,\bar{\eta}_0=Z_{\bar{\eta}}^{1/2}\bar{\eta} \;, \,\,\, \eta_0=Z_{\eta}^{1/2} \eta \;,  \\
\Xi_0&=&Z_{\Xi} \Xi\;, \,\,\, \mu^2_0=Z_{\mu^2} \mu^2 \;, \label{ra3}
\end{eqnarray}
where
\begin{eqnarray}
Z_g&&=1-\varepsilon \frac{a_0}{2} \nonumber \\
Z^{1/2}_A&&=Z^{-1}_{\Omega}=Z^{-1/2}_{\bar{c}}=Z^{-1/2}_b=Z^{1/2}_{\alpha}=1+\varepsilon d_2(\alpha) \nonumber \\
Z_\xi^{ab}&&=\delta^{ab}+\varepsilon f_{1}^{ab}(\xi,\alpha) \nonumber \\
Z_L&&=Z^{-1/2}_c=1+\varepsilon d_1(\alpha) \nonumber \\
Z_{\bar{\eta}}&&=Z_{\eta}=Z^2_{\Xi}=Z^{1/2}_{\tau}=Z_{\mathcal{J}}=1+\varepsilon a_4 \nonumber \\
Z_{m^2}&&=1+\varepsilon a_7 \nonumber \\
Z_{\mu^2}&&=1+\varepsilon (d_2-f_2(0,\alpha)) \nonumber \\
Z_{K}^{ab}&&=\delta^{ab}-\varepsilon\left(f_{1}^{ab}(\xi,\alpha)+\frac{\partial f_{1}^{kb}(\xi,\alpha)}{\partial \xi^a}\xi^{k}\right) \;. \label{ra4}
\end{eqnarray}
Notice that, as expected, the dimensionless field $\xi^a$ renormalizes in a non-linear way through the quantity $f_{1}^{ab}(\xi,\alpha)$ which is a power series in $\xi^a$.  Equations \eqref{ra1} and \eqref{ra4} establish the renormalizability of the complete action $\Sigma$, eq.\eqref{cact}, and thus of the invariant action $S$ of expression \eqref{stact}, up to a BRST exact unphysical ambiguity of the type of eq.\eqref{rp1}. As already mentioned, the explicit inclusion of such an ambiguity will be provided in Appendix \eqref{apc}.

 \section{Conclusion}
 
In this work the gauge invariant operator  $A_{\min }^{2}$, eq.\eqref{Aminn0}, and corresponding gauge invariant transverse field configuration $A^{ah}_\mu$, eq.\eqref{min0},  have been investigated in a general class of gauge fixings, eq.\eqref{gf1} and eq.\eqref{gf12},  which share similarities with 't Hooft's $R_\zeta$-gauge used in the analysis of Yang-Mills theory with spontaneous symmetry breaking.  As shown in \cite{Fiorentini:2016rwx}, a local setup can be constructed for both $A_{\min }^{2}$ and $A^{ah}_\mu$, being summarised by the local and BRST invariant action \eqref{act1}. The localization procedure makes use of an auxiliary dimensionless Stueckelberg field $\xi^a$. However, despite the presence of the field $\xi^a$ and unlike the conventional non-renormalizable Stueckelberg mass term, the present construction gives rise to a perfectly well behaved model in the ultraviolet which turns out to be renormalizable to all orders, as discussed in details in Sections \eqref{alg1} and \eqref{alg2} as well as in Appendix \eqref{apc}. In particular, the pivotal role of the transversality constraint $\partial_\mu A^{ah}_\mu =0$ has been underlined throughout the paper. It is precisely the direct implementation of this constraint in the local action \eqref{act1} which makes a substantial difference with respect to the conventional Stueckelberg theory. In fact, as pointed out in Section \eqref{St}, it removes exactly the component of the Stueckelberg propagator which gives rise to non-renormalizable ultraviolet divergences, see eq.\eqref{stp} versus eqs.\eqref{propqq}. In particular, form eqs.\eqref{propqq}, one sees that, similar to what happens in the case of  
 't Hooft's $R_\zeta$-gauge, the use of the general class of gauge fixings \eqref{gf1} and \eqref{gf12} provide a mass $\mu^2$ for the dimensionless Stueckelberg field $\xi^a$. This a welcome feature which can be effectively employed as a fully BRST invariant infrared regularization for  $\xi^a$  in explicit higher loop calculations. 
 
 \section*{Acknowledgements.}
 We thank D. Dudal, L. F. Palhares, B.W. Mintz and U. Reinosa for useful discussions. 
The Conselho Nacional de Desenvolvimento Cient\'{i}fico e
Tecnol\'{o}gico (CNPq-Brazil), the Faperj, Funda{\c{c}}{\~{a}}o de
Amparo {\`{a}} Pesquisa do Estado do Rio de Janeiro, the SR2-UERJ
and the Coordena{\c{c}}{\~{a}}o de Aperfei{\c{c}}oamento de
Pessoal de N{\'\i}vel Superior (CAPES) are gratefully acknowledged
for financial support.

\appendix

\section{Properties of the functional $f_{A}[u]$.} \label{apb}In this
 Appendix we recall some useful properties of the functional
$f_{A}[u]$
\begin{equation}
f_{A}[u]\equiv \mathrm{Tr}\int d^{4}x\,A_{\mu }^{u}A_{\mu
}^{u}=\mathrm{Tr}\int d^{4}x\left( u^{\dagger }A_{\mu
}u+\frac{i}{g}u^{\dagger }\partial _{\mu }u\right) \left(
u^{\dagger }A_{\mu }u+\frac{i}{g}u^{\dagger }\partial _{\mu
}u\right) \;. \label{fa}
\end{equation}
For a given gauge field configuration $A_{\mu }$, $f_{A}[u]$ is a functional
defined on the gauge orbit of $A_{\mu }$. Let $\mathcal{A}$ be the space of
connections $A_{\mu }^{a}$ with finite Hilbert norm $||A||$, \textit{i.e.}
\begin{equation}
||A||^{2}=\mathrm{Tr}\int d^{4}x\,A_{\mu }A{_{\mu }=}\frac{1}{2}\int
d^{4}xA_{\mu }^{a}A_{\mu }^{a}<+\infty \;, \label{norm0}
\end{equation}
and let $\mathcal{U}$ be the space of local gauge transformations $u$ such
that the Hilbert norm $||u^{\dagger }\partial {u}||$ is finite too, namely
\begin{equation}
||u^{\dagger }\partial {u}||^{2}=\mathrm{Tr}\int d^{4}x\,\left(
u^{\dagger }\partial _{\mu }u\right) \left( u^{\dagger }\partial
_{\mu }u\right) <+\infty \;. \label{norm1}
\end{equation}
\noindent The following proposition holds
\cite{Zwanziger:1990tn,Dell'Antonio:1989jn,Dell'Antonio:1991xt,vanBaal:1991zw}
\begin{itemize}
\item  \underline{Proposition}\newline
The functional $f_{A}[u]$ achieves its absolute minimum on the gauge orbit
of $A_{\mu }$.
\end{itemize}
\noindent This proposition means that there exists a $h\in
\mathcal{U}$ such that
\begin{eqnarray}
\delta f_{A}[h] &=&0\;,  \label{impl0} \\
\delta ^{2}f_{A}[h] &\ge &0\;,  \label{impl1} \\
f_{A}[h] &\le &f_{A}[u]\;,\;\;\;\;\;\;\;\forall \,u\in
\mathcal{U}\;. \label{impl2}
\end{eqnarray}
The operator $A_{\min }^{2}$ is thus given by
\begin{equation}
A_{\min }^{2}=\min_{\left\{ u\right\} }\mathrm{Tr}\int
d^{4}x\,A_{\mu }^{u}A_{\mu }^{u}=f_{A}[h]\;.  \label{a2min}
\end{equation}
Let us give a look at the two conditions (\ref{impl0}) and
(\ref{impl1}). To evaluate $\delta f_{A}[h]$ and $\delta
^{2}f_{A}[h]$ we set\footnote{The case of the gauge group $SU(N)$
is considered here.}
\begin{equation}
v=he^{ig\omega }=he^{ig\omega ^{a}T^{a}}\;,  \label{set0}
\end{equation}
\begin{equation}
\left[ T^{a},T^{b}\right] =if^{abc}\;T^c\;,\;\;\;\;\;\mathrm{Tr}\left( T^{a}T^{b}\right) =%
\frac{1}{2}\delta ^{ab}\;,  \label{st000}
\end{equation}
where $\omega $ is an infinitesimal hermitian matrix and we
compute the linear and quadratic terms of the expansion of the
functional $f_{A}[v]$ in power series of $\omega $. Let us first
obtain an expression for $A_{\mu }^{v}$
\begin{eqnarray}
A_{\mu }^{v} &=&v^{\dagger }A_{\mu }v+\frac{i}{g}v^{\dagger }\partial _{\mu
}v  \nonumber \\
&=&e^{-ig\omega }h^{\dagger }A_{\mu }he^{ig\omega }+\frac{i}{g}e^{-ig\omega
}\left( h^{\dagger }\partial _{\mu }h\right) e^{ig\omega }+\frac{i}{g}%
e^{-ig\omega }\partial _{\mu }e^{ig\omega }  \nonumber \\
&=&e^{-ig\omega }A_{\mu }^{h}e^{ig\omega }+\frac{i}{g}e^{-ig\omega }\partial
_{\mu }e^{ig\omega }\;.  \label{orbit0}
\end{eqnarray}
Expanding up to the order $\omega ^{2}$, we get
\begin{eqnarray}
A_{\mu }^{v} &=&\left( 1-ig\omega -g^{2}\frac{\omega ^{2}}{2}\right) A_{\mu
}^{h}\left( 1+ig\omega -g^{2}\frac{\omega ^{2}}{2}\right) +\frac{i}{g}\left(
1-ig\omega -g^{2}\frac{\omega ^{2}}{2}\right) \partial _{\mu }\left(
1+ig\omega -g^{2}\frac{\omega ^{2}}{2}\right)  \nonumber \\
&=&\left( 1-ig\omega -g^{2}\frac{\omega ^{2}}{2}\right) \left( A_{\mu
}^{h}+igA_{\mu }^{h}\omega -g^{2}A_{\mu }^{h}\frac{\omega ^{2}}{2}\right) +
\nonumber \\
&+&\frac{i}{g}\left( 1-ig\omega -g^{2}\frac{\omega ^{2}}{2}\right) \left(
ig\partial _{\mu }\omega -\frac{g^{2}}{2}\left( \partial _{\mu }\omega
\right) \omega -\frac{g^{2}}{2}\omega \left( \partial _{\mu }\omega \right)
\right)  \nonumber \\
&=&A_{\mu }^{h}+igA_{\mu }^{h}\omega -\frac{g^{2}}{2}A_{\mu }^{h}\omega
^{2}-ig\omega A_{\mu }^{h}+g^{2}\omega A_{\mu }^{h}\omega -\frac{g^{2}}{2}%
\omega ^{2}A_{\mu }^{h}  \nonumber \\
&+&\frac{i}{g}\left( ig\partial _{\mu }\omega -\frac{g^{2}}{2}\left(
\partial _{\mu }\omega \right) \omega -\frac{g^{2}}{2}\omega \partial _{\mu
}\omega +g^{2}\omega \partial _{\mu }\omega \right) +O(\omega ^{3})\;,
\label{ex1}
\end{eqnarray}
from which it follows
\begin{equation}
A_{\mu }^{v}=A_{\mu }^{h}+ig[A_{\mu }^{h},\omega ]+\frac{g^{2}}{2}[[\omega
,A_{\mu }^{h}],\omega ]-\partial _{\mu }\omega +i\frac{g}{2}[\omega
,\partial _{\mu }\omega ]+O(\omega ^{3})\;,  \label{A0}
\end{equation}
We now evaluate
\begin{eqnarray}
f_{A}[v] &=&\mathrm{Tr}\int d^{4}xA_{\mu }^{u}A_{\mu }^{u}  \nonumber \\
&=&\mathrm{Tr}\int d^{4}x\,\left[ \left( A_{\mu }^{h}+ig[A_{\mu }^{h},\omega ]+\frac{%
g^{2}}{2}[[\omega ,A_{\mu }^{h}],\omega ]-\partial _{\mu }\omega +i\frac{g}{2%
}[\omega ,\partial _{\mu }\omega ]+O(\omega ^{3})\right) \times
\right.
\nonumber \\
& &\left. \left( A_{\mu }^{h}+ig[A_{\mu }^{h},\omega ]+\frac{g^{2}}{2}[%
[\omega ,A_{\mu }^{h}],\omega ]-\partial _{\mu }\omega +i\frac{g}{2}[\omega
,\partial _{\mu }\omega ]+O(\omega ^{3})\right) \right]  \nonumber \\
&=&\mathrm{Tr}\int d^{4}x\,\left\{ A_{\mu }^{h}A_{\mu
}^{h}+igA_{\mu }^{h}[A_{\mu
}^{h},\omega ]+g^{2}A_{\mu }^{h}\omega {A}_{\mu }^{h}\omega -\frac{g^{2}}{2}%
A_{\mu }^{h}A_{\mu }^{h}\omega ^{2}-\frac{g^{2}}{2}A_{\mu }^{h}\omega
^{2}A_{\mu }^{h}-A_{\mu }^{h}\partial _{\mu }\omega \right.  \nonumber \\
&+&\left. i\frac{g}{2}A_{\mu }^{h}[\omega ,\partial _{\mu }\omega
]+ig[A_{\mu }^{h},\omega ]A_{\mu }^{h}-g^{2}[A_{\mu }^{h},\omega
][A_{\mu }^{h},\omega ]-ig[A_{\mu }^{h},\omega ]\partial _{\mu
}\omega +g^{2}\omega
A_{\mu }^{h}\omega A_{\mu }^{h}\right.  \nonumber \\
&-&\left. \frac{g^{2}}{2}A_{\mu }^{h}\omega ^{2}A_{\mu }^{h}-\frac{g^{2}}{2}%
\omega ^{2}A_{\mu }^{h}A_{\mu }^{h}-\partial _{\mu }\omega A_{\mu
}^{h}-ig\partial _{\mu }\omega [A_{\mu }^{h},\omega ]+\partial _{\mu }\omega
\partial _{\mu }\omega +i\frac{g}{2}[\omega ,\partial _{\mu }\omega ]A_{\mu
}^{h}\right\} +O(\omega ^{3})\nonumber\\
 &=&f_{A}[h]-\mathrm{Tr}\int d^{4}x\,\left\{
A_{\mu }^{h},\partial _{\mu }\omega
\right\} +\mathrm{Tr}\int d^{4}x\,\left( g^{2}A_{\mu }^{h}\omega A_{\mu }^{h}\omega -%
\frac{g^{2}}{2}A_{\mu }^{h}A_{\mu }^{h}\omega ^{2}-\frac{g^{2}}{2}A_{\mu
}^{h}\omega ^{2}A_{\mu }^{h}\right.  \nonumber \\
&-&\left. g^{2}[A_{\mu }^{h},\omega ][A_{\mu }^{h},\omega
]+g^{2}\omega A_{\mu }^{h}\omega A_{\mu
}^{h}-\frac{g^{2}}{2}A_{\mu }^{h}\omega ^{2}A_{\mu
}^{h}-\frac{g^{2}}{2}\omega ^{2}A_{\mu }^{h}A_{\mu }^{h}\right)
+\mathrm{Tr}\int d^{4}x\,\left( \partial _{\mu }\omega \partial
_{\mu }\omega \right.
\nonumber \\
&+&\left. i\frac{g}{2}[\omega ,\partial _{\mu }\omega ]A_{\mu
}^{h}-ig\partial _{\mu }\omega [A_{\mu }^{h},\omega ]-ig[A_{\mu
}^{h},\omega ]\partial _{\mu }\omega +i\frac{g}{2}A_{\mu
}^{h}[\omega ,\partial _{\mu }\omega ]\right) +O(\omega ^{3})
\nonumber
\end{eqnarray}
\begin{eqnarray}
&=&f_{A}[h]+2\int {d^{4}x}\,tr\left( \omega \partial _{\mu
}{A}_{\mu }^{h}\right) +\int {d^{4}x}\,tr\left\{ 2g^{2}\omega
{A}_{\mu }^{h}\omega
A_{\mu }^{h}-2g^{2}A_{\mu }^{h}A_{\mu }^{h}\omega ^{2}\right.  \nonumber \\
&-&\left. g^{2}\left( A_{\mu }^{h}\omega -\omega {A}_{\mu }^{h}\right)
\left( A_{\mu }^{h}\omega -\omega {A}_{\mu }^{h}\right) \right\} +\int {%
d^{4}x}\,tr\left( \partial _{\mu }\omega \partial _{\mu }\omega +i\frac{g}{2}%
\omega \partial _{\mu }\omega {A}_{\mu }^{h}-i\frac{g}{2}\partial _{\mu
}\omega \omega {A}_{\mu }^{h}\right.  \nonumber \\
&-&\left. ig\partial _{\mu }\omega {A}_{\mu }^{h}\omega +ig\partial _{\mu
}\omega \omega {A}_{\mu }^{h}-igA_{\mu }^{h}\omega \partial _{\mu }\omega
+ig\omega {A}_{\mu }^{h}\partial _{\mu }\omega +i\frac{g}{2}A_{\mu
}^{h}\omega \partial _{\mu }\omega -i\frac{g}{2}A_{\mu }^{h}\partial _{\mu
}\omega \omega \right) +O(\omega ^{3})  \nonumber \\
&=&f_{A}[h]+2\mathrm{Tr}\int d^{4}x\left( \,\omega \partial _{\mu
}A_{\mu }^{h}\right) +\mathrm{Tr}\int d^{4}x\,\left( \partial
_{\mu }\omega
\partial _{\mu }\omega +ig\omega \partial _{\mu }\omega {A}_{\mu
}^{h}-ig\partial _{\mu
}\omega \omega {A}_{\mu }^{h}\right.  \nonumber \\
&-&\left. 2ig\partial _{\mu }\omega A_{\mu }^{h}\omega +2ig\partial
_{\mu }\omega \omega A_{\mu }^{h}\right) +O(\omega ^{3})\;.
\end{eqnarray}
Thus
\begin{eqnarray}
f_{A}[v] &=&f_{A}[h]+2\mathrm{Tr}\int d^{4}x\,\left( \omega
\partial _{\mu }A_{\mu }^{h}\right) +\mathrm{Tr}\int d^{4}x\,\left(
\partial _{\mu }\omega \partial _{\mu }\omega +ig\omega \partial
_{\mu }\omega A_{\mu }^{h}-ig\partial _{\mu
}\omega \omega A_{\mu }^{h}\right.  \nonumber \\
&-&\left. ig\left( \partial _{\mu }\omega \right) A_{\mu }^{h}\omega
+ig\left( \partial _{\mu }\omega \right) \omega A_{\mu }^{h}\right)
+O(\omega ^{3})  \nonumber \\
&=&f_{A}[h]+2\mathrm{Tr}\int d^{4}x\,\left( \omega \partial _{\mu
}A_{\mu }^{h}\right) +\mathrm{Tr}\int d^{4}x\,\left\{ \partial
_{\mu }\omega \left(
\partial _{\mu }\omega -ig\left[ A_{\mu }^{h},\omega \right]
\right) \right\} +O(\omega ^{3})\;. \nonumber  \\ \label{f1}
\end{eqnarray}
Finally
\begin{equation}
f_{A}[v]=f_{A}[h]+2\mathrm{Tr}\int d^{4}x\,\left( \omega \partial
_{\mu }A_{\mu }^{h}\right) -\mathrm{Tr}\int d^{4}x\,\omega
\partial _{\mu }D_{\mu }(A^{h})\omega +O(\omega ^{3})\;,
\label{func2}
\end{equation}
so that
\begin{eqnarray}
\delta f_{A}[h] &=&0\;\;\;\Rightarrow \;\;\;\partial _{\mu }A_{\mu
}^{h}\;=\;0\;,  \nonumber \\
\delta ^{2}f_{A}[h] &>&0\;\;\;\Rightarrow \;\;\;-\partial _{\mu }D{_{\mu }(}%
A^{h}{)}\;>\;0\;.  \label{func3}
\end{eqnarray}
We see therefore that the set of field configurations fulfilling conditions (%
\ref{func3}), \textit{i.e.} defining relative minima of the functional $%
f_{A}[u]$, belong to the so called Gribov region $\Omega $, which is defined
as
\begin{equation}
\Omega =\left.\{A_{\mu }\right|\partial _{\mu }A_{\mu
}=0\;\mathrm{and}\;-\partial _{\mu }D_{\mu }(A)>0\}\;.
\label{gribov0}
\end{equation}
Let us proceed now by showing that the transversality condition,
$\partial
_{\mu }A_{\mu }^{h}=0$, can be solved for $h=h(A)$ as a power series in $%
A_{\mu }$. We start from
\begin{equation}
A_{\mu }^{h}=h^{\dagger }A_{\mu }h+\frac{i}{g}h^{\dagger }\partial _{\mu
}h\;,  \label{Ah0}
\end{equation}
with
\begin{equation}
h=e^{ig\phi }=e^{ig\phi ^{a}T^{a}}\;.  \label{h0}
\end{equation}
Let us expand $h$ in powers of $\phi $
\begin{equation}
h=1+ig\phi -\frac{g^{2}}{2}\phi ^{2}+O(\phi ^{3})\;.  \label{hh1}
\end{equation}
From equation (\ref{Ah0}) we have
\begin{equation}
A_{\mu }^{h}=A_{\mu }+ig[A_{\mu },\phi ]+g^{2}\phi A_{\mu }\phi -\frac{g^{2}%
}{2}A_{\mu }\phi ^{2}-\frac{g^{2}}{2}\phi ^{2}A_{\mu }-\partial _{\mu }\phi
+i\frac{g}{2}[\phi ,\partial _{\mu }]+O(\phi ^{3})\;.  \label{A1}
\end{equation}
Thus, condition $\partial _{\mu }A_{\mu }^{h}=0$, gives
\begin{eqnarray}
\partial ^{2}\phi &=&\partial _{\mu }A+ig[\partial _{\mu }A_{\mu },\phi
]+ig[A_{\mu },\partial _{\mu }\phi ]+g^{2}\partial _{\mu }\phi A_{\mu }\phi
+g^{2}\phi \partial _{\mu }A_{\mu }\phi +g^{2}\phi A_{\mu }\partial _{\mu
}\phi   \nonumber \\
&-&\frac{g^{2}}{2}\partial _{\mu }A_{\mu }\phi ^{2}-\frac{g^{2}}{2}A_{\mu
}\partial _{\mu }\phi \phi -\frac{g^{2}}{2}A_{\mu }\phi \partial _{\mu }\phi
-\frac{g^{2}}{2}\partial _{\mu }\phi \phi A_{\mu }-\frac{g^{2}}{2}\phi
\partial _{\mu }\phi A_{\mu }-\frac{g^{2}}{2}\phi ^{2}\partial _{\mu }A_{\mu
}  \nonumber \\
&+&i\frac{g}{2}[\phi ,\partial ^{2}\phi ]+O(\phi ^{3})\;.  \label{hh2}
\end{eqnarray}
This equation can be solved iteratively for $\phi $ as a power series in $%
A_{\mu }$, namely
\begin{equation}
\phi =\frac{1}{\partial ^{2}}\partial _{\mu }A_{\mu }+i\frac{g}{\partial ^{2}%
}\left[ \partial A,\frac{\partial A}{\partial ^{2}}\right] +i\frac{g}{%
\partial ^{2}}\left[ A_{\mu },\partial _{\mu }\frac{\partial A}{\partial ^{2}%
}\right] +\frac{i}{2}\frac{g}{\partial ^{2}}\left[ \frac{\partial A}{%
\partial ^{2}},\partial A\right] +O(A^{3})\;,  \label{phi0}
\end{equation}
so that
\begin{eqnarray}
A_{\mu }^{h} &=&A_{\mu }-\frac{1}{\partial ^{2}}\partial _{\mu }\partial A-ig%
\frac{\partial _{\mu }}{\partial ^{2}}\left[ A_{\nu },\partial _{\nu }\frac{%
\partial A}{\partial ^{2}}\right] -i\frac{g}{2}\frac{\partial _{\mu }}{%
\partial ^{2}}\left[ \partial A,\frac{1}{\partial ^{2}}\partial A\right]
\nonumber \\
&+&ig\left[ A_{\mu },\frac{1}{\partial ^{2}}\partial A\right] +i\frac{g}{2}%
\left[ \frac{1}{\partial ^{2}}\partial A,\frac{\partial _{\mu }}{\partial
^{2}}\partial A\right] +O(A^{3})\;.  \label{minn2}
\end{eqnarray}
Expression (\ref{minn2}) can be written in a more useful way,
given in eq.(\ref{min0}). In fact
\begin{eqnarray}
A_{\mu }^{h} &=&\left( \delta _{\mu \nu }-\frac{\partial _{\mu }\partial
_{\nu }}{\partial ^{2}}\right) \left( A_{\nu }-ig\left[ \frac{1}{\partial
^{2}}\partial A,A_{\nu }\right] +\frac{ig}{2}\left[ \frac{1}{\partial ^{2}}%
\partial A,\partial _{\nu }\frac{1}{\partial ^{2}}\partial A\right] \right)
+O(A^{3})  \nonumber \\
&=&A_{\mu }-ig\left[ \frac{1}{\partial ^{2}}\partial A,A_{\mu }\right] +%
\frac{ig}{2}\left[ \frac{1}{\partial ^{2}}\partial A,\partial _{\mu }\frac{1%
}{\partial ^{2}}\partial A\right] -\frac{\partial _{\mu }}{\partial ^{2}}%
\partial A+ig\frac{\partial _{\mu }}{\partial ^{2}}\partial _{\nu }\left[
\frac{1}{\partial ^{2}}\partial A,A_{\nu }\right]   \nonumber \\
&-&i\frac{g}{2}\frac{\partial _{\mu }}{\partial ^{2}}\partial _{\nu }\left[
\frac{\partial A}{\partial ^{2}},\frac{\partial _{\nu }}{\partial ^{2}}%
\partial A\right] +O(A^{3})  \nonumber \\
&=&A_{\mu }-\frac{\partial _{\mu }}{\partial ^{2}}\partial A+ig\left[ A_{\mu
},\frac{1}{\partial ^{2}}\partial A\right] +\frac{ig}{2}\left[ \frac{1}{%
\partial ^{2}}\partial A,\partial _{\mu }\frac{1}{\partial ^{2}}\partial
A\right] +ig\frac{\partial _{\mu }}{\partial ^{2}}\left[ \frac{\partial
_{\nu }}{\partial ^{2}}\partial A,A_{\nu }\right]   \nonumber \\
&+&i\frac{g}{2}\frac{\partial _{\mu }}{\partial ^{2}}\left[ \frac{\partial A%
}{\partial ^{2}},\partial A\right] +O(A^{3})  \label{hhh3}
\end{eqnarray}
which is precisely expression (\ref{minn2}). The transverse field
given in eq.(\ref {min0}) enjoys the property of being gauge
invariant order by order in the
coupling constant $g$. Let us work out the transformation properties of $%
\phi _{\nu }$ under a gauge transformation
\begin{equation}
\delta A_{\mu }=-\partial _{\mu }\omega +ig[A_{\mu },\omega ]\;.
\label{gauge3}
\end{equation}
We have, up to the order $O(g^{2})$,
\begin{eqnarray}
\delta \phi _{\nu } &=&-\partial _{\nu }\omega +ig\left[ \frac{1}{\partial
^{2}}\partial A,\partial _{\nu }\omega \right] -i\frac{g}{2}\left[ \omega
,\partial _{\nu }\frac{1}{\partial ^{2}}\partial A\right] -i\frac{g}{2}%
\left[ \frac{\partial A}{\partial ^{2}},\partial _{\nu }\omega \right]
+O(g^{2})  \nonumber \\
&=&-\partial _{\nu }\omega +i\frac{g}{2}\left[ \frac{1}{\partial ^{2}}%
\partial A,\partial _{\nu }\omega \right] +i\frac{g}{2}\left[ \partial _{\nu
}\frac{1}{\partial ^{2}}\partial A,\omega \right] +O(g^{2})\;.  \label{gg2}
\end{eqnarray}
Therefore
\begin{equation}
\delta \phi _{\nu }=-\partial _{\nu }\left( \omega -i\frac{g}{2}\left[ \frac{%
\partial A}{\partial ^{2}},\omega \right] \right) +O(g^{2})\;,  \label{phi1}
\end{equation}
from which the gauge invariance of $A_{\mu }^{h}$ is established.\newline
\newline
Finally, let us work out the expression of $A_{\mathrm{min}}^{2}$ as a power
series in $A_{\mu }$.
\begin{eqnarray}
A_{\mathrm{min}}^{2} &=&\mathrm{Tr}\int d^{4}x\,A_{\mu }^{h}A_{\mu }^{h}  \nonumber \\
&=&\mathrm{Tr}\int d^{4}x\,\left[ \phi _{\mu }\left( \delta _{\mu \nu }-\frac{%
\partial _{\mu }\partial _{\nu }}{\partial ^{2}}\right) \phi _{\nu }\right]
\nonumber \\
&=&\mathrm{Tr}\int d^{4}x\,\left[ \left( A_{\mu }-ig\left[ \frac{1}{\partial ^{2}}%
\partial A,A_{\mu }\right] +\frac{ig}{2}\left[ \frac{1}{\partial ^{2}}%
\partial A,\partial _{\mu }\frac{1}{\partial ^{2}}\partial A\right] \right)
\times \right.  \nonumber \\
&&\left.  \left( \delta _{\mu \nu }-\frac{\partial _{\mu }\partial
_{\nu }}{\partial ^{2}}\right) \left( A_{\nu }-ig\left[
\frac{1}{\partial
^{2}}\partial A,A_{\nu }\right] +\frac{ig}{2}\left[ \frac{1}{\partial ^{2}}%
\partial A,\partial _{\nu }\frac{1}{\partial ^{2}}\partial A\right] \right)
\right]  \nonumber \\
&=&\frac{1}{2}\int d^{4}x\left[ A_{\mu }^{a}\left( \delta _{\mu \nu }-\frac{%
\partial _{\mu }\partial _{\nu }}{\partial ^{2}}\right) A_{\nu
}^{a}-2gf^{abc}\frac{\partial _{\nu }\partial A^{a}}{\partial ^{2}}\frac{%
\partial A^{b}}{\partial ^{2}}A_{\nu }^{c}-gf^{abc}A_{\nu }^{a}\frac{%
\partial A^{b}}{\partial ^{2}}\frac{\partial _{\nu }\partial A^{c}}{\partial
^{2}}\right] +O(A^{4})\;.  \nonumber \\
&&  \label{a2em}
\end{eqnarray}
We conclude this Appendix by noting that, due to gauge invariance, $A_{%
\mathrm{min}}^{2}$ can be rewritten in a manifestly invariant way
in terms of $F_{\mu \nu }$ and the covariant derivative $D_{\mu }$
\cite{Zwanziger:1990tn}.

\section{A generalised Slavnov-Taylor identity}  \label{apc}
In this Appendix we derive the Ward identities for the generalised gauge fixing  of eq.\eqref{gf12}. Since 
the quantity $\omega^a(\xi)$ is now a composite operator, {\it i.e.} a product of fields at the same space-time point, 
we need to define $\omega^a(\xi)$ by introducing it into the starting action though a suitable external source. In order to maintain BRST invariance, we 
make use of a BRST doublet of external sources $(Q^a, R^a)$, of dimension four and ghost number $(-1,0)$, 
\begin{equation}
sQ^a = R^a \;, \qquad sR^a = 0 \;, \label{qr} 
\end{equation} 
and introduce the term 
\begin{equation} 
\int d^4x \; s \left( Q^a \omega^a(\xi) \right) = \int d^4 x \left( R^a \omega^a(\xi) - Q^a \frac{\partial \omega^a}{\partial \xi^c} g^{cd}(\xi) c^d \right)  \;. \label{qr1} 
\end{equation}
We start thus with the complete classical action $\Sigma$ given now by 
\begin{eqnarray} 
\Sigma &=&  S_{inv} + \int d^4x \left( {\cal J}^a_\mu A^{ah}_\mu  + \Xi^a_\mu D^{ab}_\mu(A^h) \eta^b \right)  \nonumber \\
&+&   \int d^4x \left( i b^a \partial_\mu A^a_\mu +  \frac{\alpha}{2} b^a b^a 
- i M^{ab} b^a \omega^b(\xi) - N^{ab} {\bar c}^a \omega^b(\xi) + {\bar c}^a \partial_\mu D^{ab}_\mu c^b + M^{ab} {\bar c}^a \frac{\partial \omega^{b}(\xi)}{\partial \xi^c} g^{cd}(\xi) c^d  \right)  \nonumber \\
&+& \int d^4x \left( -\Omega^a_\mu D^{ab}_\mu c^b + L^a \frac{g f^{abc}}{2} c^b c^c + K^a g^{ab}(\xi) c^d + R^a \omega^a(\xi) - Q^a \frac{\partial \omega^a}{\partial \xi^c} g^{cd}(\xi) c^d \right) \;,  \label{ca1}
\end{eqnarray} 
with $S_{inv}$ given by expression \eqref{act1}. \\\\The action $\Sigma$, eq.\eqref{ca1},  obeys the following Ward identities:
\begin{itemize} 
\item the Slavnov-Taylor identity
\begin{equation}
\int d^4x \left( \frac{\delta \Sigma}{\delta A^a_\mu} \frac{\delta \Sigma}{\delta \Omega^a_\mu}  +  \frac{\delta \Sigma}{\delta c^a} \frac{\delta \Sigma}{\delta L^a}  
+ \frac{\delta \Sigma}{\delta \xi^a} \frac{\delta \Sigma}{\delta K^a} + ib^a \frac{\delta \Sigma}{\delta {\bar c}^a} + N^{ab} \frac{\delta \Sigma}{\delta M^{ab}}
+ R^{a} \frac{\delta \Sigma}{\delta Q^{a}}\right)   = 0 \;, \label{stg}
\end{equation}
\item the equation of motion of the Lagrange multiplier $b^a$
\begin{equation} 
\frac{\delta \Sigma}{\delta b^a} = i \partial_\mu A^a_\mu + \alpha b^a - i M^{ab} \frac{\delta \Sigma}{\delta R^b} \;, \label{bg1} 
\end{equation}  
\item the anti-ghost equation 
\begin{equation}
\frac{\delta \Sigma}{\delta {\bar c}^a} + \partial_\mu \frac{\delta \Sigma}{\delta \Omega^a_\mu}  + M^{ab} \frac{\delta \Sigma}{\delta Q^b}  - N^{ab} \frac{\delta \Sigma}{\delta R^b} = 0 \;, \label{ghg1}
\end{equation}
\item the equation of  $\tau^{a}$
\begin{eqnarray}
\frac{\delta \Sigma}{\delta\tau^{a}}-\partial_{\mu}\frac{\delta \Sigma}{\delta\mathcal{J}_{\mu}^{a}} & = & 0 \;, \label{ap5} 
\end{eqnarray}
\item the equation of the ghost $\eta^{a}$
\begin{eqnarray}
\int d^{4}x\left(\frac{\delta \Sigma}{\delta\eta^{a}}+gf^{abc}\bar{\eta}^{b}\frac{\delta \Sigma}{\delta\tau^{c}}+gf^{abc}\Xi^{b}\frac{\delta \Sigma}{\delta\mathcal{J}_{\mu}^{c}}\right) & = & 0  \;, \label{ap6}
\end{eqnarray}
\item the equation of the antighost ${\bar \eta}^a$
\begin{eqnarray}
\frac{\delta \Sigma}{\delta\bar{\eta}^{a}}-\partial_{\mu}\frac{\delta \Sigma}{\delta\Xi_{\mu}^{a}} & = & 0 \;. \label{ap7}
\end{eqnarray}
\end{itemize}
These Ward identities can be employed for the analysis of the algebraic renormalization when the generalised function $\omega^a(\xi)$ is explicitly present in the 
gauge-fixing. In this case, the general counterterm will be reabsorbed through a renormalization of $\omega^a(\xi)$, corresponding to a renormalization of the infinite set of unphysical gauge parameters $(a_1^{abc}, a_2^{abcd}, a_3^{abcde}, ..)$  of expression \eqref{rp}. \\\\Repeating the lengthy discussion of the previous sections, for the most general local invariant counterterm we find now

\begin{eqnarray}
\Sigma^{ct} & = & \int d^4 x \; \bigg\{ -a_{0}g^2\frac{\partial \Sigma}{\partial g^2}+d_{2}\left(\alpha\right)2\alpha\frac{\partial \Sigma}{\partial\alpha}+a_{7} m^{2}\frac{\partial \Sigma}{\partial m^{2} \Bigr. \nonumber }\\
 &  & +a_{4}\left(\tau^{a}\frac{\delta \Sigma}{\delta\tau^{a}}+\mathcal{J}_{\mu}^{a}\frac{\delta \Sigma}{\delta\mathcal{J}_{\mu}^{a}}+\frac{1}{2}\bar{\eta}^{a}\frac{\delta \Sigma}{\delta\bar{\eta}^{a}}+\frac{1}{2}\eta^{a}\frac{\delta \Sigma}{\delta \eta^{a}}+\frac{1}{2}\Xi_{\mu}^{a}\frac{\delta \Sigma}{\delta\Xi_{\mu}^{a}}\right) \nonumber \\
 & &+d_{2}\left(\alpha\right)A_{\mu}^{a}\frac{\delta \Sigma}{\delta A_{\mu}^{a}}-d_{2}\left(\alpha\right)\Omega_{\mu}^{a}\frac{\delta \Sigma}{\delta\Omega_{\mu}^{a}}-d_{1}\left(\alpha\right)c^{a}\frac{\delta \Sigma}{\delta c^{a}}+d_{1}\left(\alpha\right)L^{a}\frac{\delta \Sigma}{\delta L^{a}} \nonumber \\
 &&+f_{1}^{ab}(\xi,\alpha)\xi^{a}\frac{\delta \Sigma}{\delta \xi^b }-\left(f_1^{ab}(\xi,\alpha)+\frac{\partial f_{1}^{kb}(\xi,\alpha)}{\partial\xi^a}\xi^k\right)K^b\frac{\delta \Sigma}{\delta K^a } \nonumber \\
& &-d_2(\alpha) \bar{c}\frac{\delta \Sigma}{\delta \bar{c}}+\left(d_2(\alpha)-f_2(0,\alpha)\right) M^{ab} \frac{\delta \Sigma}{\delta M^{ab}}+\left(d_2(\alpha)-f_2(0,\alpha)\right)N^{ab}\frac{\delta \Sigma}{\delta N^{ab}}\nonumber \\
& &-d_2(\alpha) b^a\frac{\delta \Sigma}{\delta b^a}-f_2(0,\alpha) Q^a \frac{\delta \Sigma}{\delta Q^a}-f_2(0,\alpha)R^a\frac{\delta \Sigma}{\delta R^a}\nonumber \\
& &+ \big[ \left(f_2(0,\alpha)a_1^{abc}+\tilde a_1^{abc}\right) \frac{\delta \Sigma}{\delta a_1^{abc}}+\left(f_2(0,\alpha)a_2^{abcd}+\tilde{a}_2^{abcd}\right)\frac{\delta \Sigma}{\delta a_2^{abcd}} \nonumber \\
& &\Bigl. +\left(f_2(0,\alpha)a_3^{abcde}+\tilde{a}_3^{abcde}\right)\frac{\delta \Sigma}{\delta a_3^{abcde}}+... \big] \bigg\} \;,   \label{cctt}
\end{eqnarray}
where the dots $ ... $ denote the reamaining, infinite set, of terms of the kind  
\begin{equation} 
\sum_{j} \left(f_2(0,\alpha)a_j^{abcde...}+\tilde{a}_j^{abcde...}\right)\frac{\delta \Sigma}{\delta a_j^{abcde...}}  \;, \qquad \ j=4, ..., \infty  \;, \label{jin}
\end{equation} 
The counterterm $\Sigma^{ct}$ in eq.\eqref{cctt}  can be rewritten  as 

\begin{eqnarray}
\Sigma^{ct} = \mathcal{R}\Sigma \;,  \label{rpp}
\end{eqnarray}
with
\begin{eqnarray}
\mathcal{R}&=& -a_{0}g^2\frac{\partial}{\partial g^2}+d_{2}\left(\alpha\right)2\alpha\frac{\partial}{\partial\alpha}+a_{7} m^{2}\frac{\partial }{\partial m^{2}} \nonumber \\
 &  & +\int d^4x \,\bigg\{a_{4}\left(\tau^{a}\frac{\delta}{\delta\tau^{a}}+\mathcal{J}_{\mu}^{a}\frac{\delta }{\delta\mathcal{J}_{\mu}^{a}}+\frac{1}{2}\bar{\eta}^{a}\frac{\delta }{\delta\bar{\eta}^{a}}+\frac{1}{2}\eta^{a}\frac{\delta }{\delta \eta^{a}}+\frac{1}{2}\Xi_{\mu}^{a}\frac{\delta }{\delta\Xi_{\mu}^{a}}\right) \nonumber \\
 & &+d_{2}\left(\alpha\right)A_{\mu}^{a}\frac{\delta}{\delta A_{\mu}^{a}}-d_{2}\left(\alpha\right)\Omega_{\mu}^{a}\frac{\delta}{\delta\Omega_{\mu}^{a}}-d_{1}\left(\alpha\right)c^{a}\frac{\delta}{\delta c^{a}}+d_{1}\left(\alpha\right)L^{a}\frac{\delta}{\delta L^{a}} \nonumber \\
 &&+f_{1}^{ab}(\xi,\alpha)\xi^{a}\frac{\delta }{\delta \xi^b }-\left(f_1^{ab}(\xi,\alpha)+\frac{\partial f_{1}^{kb}(\xi,\alpha)}{\partial\xi^a}\xi^k\right)K^b\frac{\delta }{\delta K^a } \nonumber \\
& &-d_2(\alpha) \bar{c}\frac{\delta}{\delta \bar{c}}+\left(d_2(\alpha)-f_2(0,\alpha)\right) M^{ab} \frac{\delta }{\delta M^{ab}}+\left(d_2(\alpha)-f_2(0,\alpha)\right)N^{ab}\frac{\delta}{\delta N^{ab}} \nonumber \\
& &-d_2(\alpha) b^a\frac{\delta}{\delta b^a}-f_2(0,\alpha) Q^a \frac{\delta }{\delta Q^a}-f_2(0,\alpha)R^a\frac{\delta}{\delta R^a} \nonumber \\
& &+ \big[ \left(f_2(0,\alpha)a_1^{abc}+\tilde a_1^{abc}\right) \frac{\delta}{\delta a_1^{abc}}+\left(f_2(0,\alpha)a_2^{abcd}+\tilde{a}_2^{abcd}\right)\frac{\delta }{\delta a_2^{abcd}} \nonumber \\
& &+\left(f_2(0,\alpha)a_3^{abcde}+\tilde{a}_3^{abcde}\right)\frac{\delta }{\delta a_3^{abcde}}+... \big]\bigg\} \;. \label{rpp1}
\end{eqnarray}
For the renormalization factors, we have now 
\begin{equation} 
 \Sigma(\Phi) +  \varepsilon  \Sigma^{ct}(\Phi)   =   \Sigma(\Phi) + \varepsilon \mathcal{R}\Sigma(\Phi) =  \Sigma(\Phi_0) + O(\varepsilon^2) \;, \label{rpp2}
 \end{equation}
with
\begin{eqnarray}
\Phi_0=Z_{\Phi} \Phi= (1+\varepsilon \mathcal{R})\Phi + O(\varepsilon^2) \;. \label{rpp3} 
\end{eqnarray}
where
\begin{eqnarray}
A_0&=&Z_A^{1/2} A_{\mu}\;, \,\,\, b_0=Z_b^{1/2}\;, \,\,\, c_0=Z_c^{1/2}c\;, \,\,\, \bar{c}_0=Z_{\bar{c}}^{1/2}\bar{c}\;, \nonumber  \\
\xi_0^a& = & Z^{ab}_{\xi}(\xi)\xi^b, \,\,\, \tau_0=Z_{\tau}^{1/2} \tau \;, 
\Omega_0=Z_{\Omega} \Omega \;, \,\, \,L_0=Z_L L\;,  \nonumber \\
 \,\,\, K_0^a&=&Z_K^{ab} (\xi) K^b\;, \,\,\, m_0^2=Z_{m^2} m^2\;, \,\,\,\mathcal{J}_0=Z_{\mathcal{J}}\mathcal{J} \;, \nonumber \\
g_0&=&Z_g\;, \,\,\, \alpha_0=Z_{\alpha} \alpha\;, \,\,\,\bar{\eta}_0=Z_{\bar{\eta}}^{1/2}\bar{\eta}\;, \,\,\, \eta_0=Z_{\eta}^{1/2} \eta\;, \nonumber \\
 \Xi_0&=& Z_{\Xi} \Xi\;, \,\,\, M_0=Z_M M\;,  \nonumber \\
N_0&=&Z_N N, \,\,\, Q_0=Z_Q Q, \,\,\, R_0= Z_R R \;, \label{rpp4} 
\end{eqnarray}
and
\begin{eqnarray}
Z_g&&=1-\varepsilon \frac{a_0}{2}  \nonumber \\
Z^{1/2}_A&&=Z^{-1}_{\Omega}=Z^{-1/2}_{\bar{c}}=Z^{-1/2}_b=Z^{1/2}_{\alpha}=1+\varepsilon d_2(\alpha) \nonumber \\
Z_\xi^{ab}&&=\delta^{ab}+\varepsilon f_{1}^{ab}(\xi,\alpha) \nonumber \\
Z_L&&=Z^{-1/2}_c=1+\varepsilon d_1(\alpha) \nonumber \\
Z_{\bar{\eta}}&&=Z_{\eta}=Z^2_{\Xi}=Z^{1/2}_{\tau}=Z_{\mathcal{J}}=1+\varepsilon a_4 \nonumber \\
Z_{m^2}&&=1+\varepsilon a_7 \nonumber \\
Z_{M}&&=Z_{N}=1+\varepsilon (d_2-f_2(0,\alpha)) \nonumber \\
Z_Q&&=Z_R=1-\varepsilon (f_2(0,\alpha)) \nonumber \\
Z_{K}^{ab}&&=\delta^{ab}-\varepsilon\left(f_{1}^{ab}(\xi,\alpha)+\frac{\partial f_{1}^{kb}(\xi,\alpha)}{\partial \xi^a}\xi^{k}\right) \;,  \label{rpp5}
\end{eqnarray}
with the addition of a multiplicative renormalization of the infinite set of gauge parameters $(a_1^{abc}, a_2^{abcd}, a_3^{abcde}, ..)$  of expression \eqref{rp}, namely 
\begin{eqnarray}
(a_1^{abc})_0&=&(1+\varepsilon f_2(0,\alpha))a_1^{abc}+\varepsilon\tilde{a}_1^{abc} \nonumber \\
(a_2^{abcd})_0&=&(1+\varepsilon f_2(0,\alpha))a_2^{abcd}+\varepsilon\tilde{a}_2^{abcd}  \nonumber \\
(a_3^{abcde})_0&=&(1+\varepsilon f_2(0,\alpha))a_3^{abcde}+\varepsilon\tilde{a}_3^{abcde} \nonumber \\
&...& \;.  \label{rpp6} 
\end{eqnarray}
Equations \eqref{rpp5} and  \eqref{rpp6}  show that the inclusion of the ambiguity $\omega^a(\xi)$ in the generalised gauge fixing of eq.\eqref{gf12} gives rise to a standard renormalization of the fields, parameters and sources. Clearly, from eq.\eqref{rpp6}  one sees that the renormalization of $\omega^a(\xi)$ itself is now encoded into a multiplicative renormalization of the infinite set of the unphysical gauge parameters $(a_1^{abc}, a_2^{abcd}, a_3^{abcde}, ..)$.

\end{document}